\PassOptionsToPackage{pdfpagelabels=false}{hyperref}

\documentclass[fleqn,usenatbib]{mnras}
\usepackage{newtxtext,newtxmath}

\usepackage[T1]{fontenc}
\usepackage{ae,aecompl}
\usepackage{graphicx,epsfig}	
\usepackage{amsmath}	
\usepackage{amssymb}	
\usepackage{booktabs} 
\usepackage[dvipsnames]{xcolor}
\usepackage{color, colortbl}

\title[Origins of ultra-diffuse galaxies in the Coma cluster - II]{Origins of ultra-diffuse galaxies in the Coma cluster - II. Constraints from their stellar populations}
\author[A. Ferr\'e-Mateu et al.]{Anna Ferr\'e-Mateu$^{1}$\thanks{E-mail: aferremateu@swin.edu.au (AFM)}, Adebusola  Alabi$^{2}$, Duncan A. Forbes$^{1}$, Aaron J. Romanowsky$^{3,2}$, 
\newauthor  Jean Brodie$^{2}$, Viraj Pandya$^{2}$, Ignacio Mart\'in-Navarro$^{2}$, Sabine Bellstedt$^{1}$, Asher Wasserman$^{2}$, 
\newauthor  Maria B. Stone$^{3,4}$ and Nobuhiro Okabe$^{5,6,7}$\\
$^{1}$ Centre for Astrophysics \& Supercomputing, Swinburne University of Technology, Hawthorn VIC 3122, Australia\\
$^{2}$ University of California Observatories, 1156 High St., Santa Cruz, CA 95064, USA\\
$^{3}$ Department of Physics and Astronomy, San Jos\'e State University, San Jose, CA 95192, USA \\	
$^{4}$ Department of Physics and Astronomy, University of Turku, FI-20014, Finland\\
$^{5}$ Department of Physical Science, Hiroshima University, 1-3-1, Kagamiyama, Higashi-Hiroshima, Hiroshima 739-8526, Japan\\
$^{6}$ Hiroshima Astrophysical Science Center, Hiroshima University, 1-3-1, Kagamiyama, Higashi-Hiroshima, Hiroshima 739-8526, Japan\\
$^{7}$ Core Research for Energetic Universe, Hiroshima University, 1-3-1, Kagamiyama, Higashi-Hiroshima, Hiroshima 739-8526, Japan
}
\date{Accepted 2018 June 9. Received 2018 May 18; in original form 2018 January 26}

\pubyear{xxxx}

\begin{document}
\label{firstpage}
\pagerange{\pageref{firstpage}-\pageref{lastpage}}
\maketitle

\begin{abstract}
In this second paper of the series we study, with new Keck/DEIMOS spectra, the stellar populations of seven spectroscopically confirmed ultra-diffuse galaxies (UDGs) in the Coma cluster. We find intermediate to old ages ($\sim$\,7\,Gyr), low metallicities ([Z/H]$\sim$\,--\,0.7\,dex) and mostly super-solar abundance patterns ([Mg/Fe] $\sim$\,0.13\,dex). These properties are similar to those of low-luminosity (dwarf) galaxies inhabiting the same area in the cluster and are mostly consistent with being the continuity of the stellar mass scaling relations of more massive galaxies. These UDGs' star formation histories imply a relatively recent infall into the Coma cluster, consistent with the theoretical predictions for a dwarf-like origin. However, considering the scatter in the resulting properties and including other UDGs in Coma, together with the results from the velocity phase-space study of the Paper I in this series, a mixed-bag of origins is needed to explain the nature of all UDGs. Our results thus reinforce a scenario in which many UDGs are field dwarfs that become quenched through their later infall onto cluster environments, whereas some UDGs could be be genuine primordial galaxies that failed to develop due to an early quenching phase. The unknown proportion of dwarf-like to primordial-like UDGs leaves the enigma of the nature of UDGs still open.
\end{abstract}

\begin{keywords}
galaxies: evolution - galaxies: formation - galaxies: kinematics and dynamics - galaxies: stellar content 
\end{keywords}

\section{Introduction}
Although low-surface brightness galaxies were discovered many decades ago (e.g. \citealt{Impey1988}; \citealt{Dalcanton1997}), the recent finding of a large number of them in the Coma cluster (e.g. \citealt{vanDokkum2015a}) has refuelled the interest by the scientific community for this intriguing class of galaxies. Relabelled as ultra-diffuse galaxies (UDGs), they have been reported to exist across a wide range of environments. Large numbers of them have been reported in clusters (e.g. \citealt{Koda2015}, \citealt{Munoz2015}; \citealt{Mihos2015}; \citealt{Yagi2016}; \citealt{vanderBurg2016}; \citealt{Roman2017a}; \citealt{Janssens2017}; \citealt{Lee2017}; \citealt{Venhola2017}; \citealt{Wittmann2017}), but others also in groups (e.g. \citealt{Makarov2015}; \citealt{Merritt2016}; \citealt{Roman2017b}; \citealt{Trujillo2017}; \citealt{vanderBurg2017}, \citealt{Shi2017}) and occasionally in the field (e.g. \citealt{MartinezDelgado2016}; \citealt{Bellazzini2017}; \citealt{Leisman2017}; \citealt{Papastergis2017}). 

UDGs share similar luminosities and stellar masses to dwarf galaxies (L$\mathrm{_V}\sim$10$^{8}$\,L$_{\odot}$; M$_{\ast}\sim$10$^{7}$-\,10$^{8}$M$_{\odot}$), but are as large as giant ellipticals ($\mathrm{R_e}\sim$1.5--\,4.6\,kpc) with typically prolate-spheroidal shapes \citep{Burkert2017}. Using the number of globular clusters and their relation with the stellar and halo galaxy masses, some UDGs have been found to contain large amounts of dark matter (e.g. \citealt{Beasley2016}; \citealt{vanDokkum2016}; \citealt{vanDokkum2017}), whereas others have been found compatible with the halos of dwarfs galaxies (e.g. \citealt{Amorisco2018}; \citealt{Toloba2018}). There has even been one UDG reported to completely lack dark matter (\citealt{vanDokkum2018a} and \citealt{vanDokkum2018b}; but see e.g. \citealt{Laporte2018}; \citealt{Ogiya2018} with possible explanations for this effect). In any case, the variety of properties and the extravaganza of some of them has prompted to a myriad of discussions debating the origins and formation scenarios for UDGs.

So far, various pathways of formation have been proposed for UDGs. One such pathway is that they are `failed' galaxies (e.g. \citealt{vanDokkum2015b}; \citealt{Beasley2016}; \citealt{BeasleyT2016}; \citealt{Peng2016}; \citealt{vanDokkum2016}). Under this scenario, the assumption is that the star formation of the primordial halo was truncated at very early epochs ($z>2$), most likely produced by the infall of such undeveloped halos into the cluster environment (\citealt{Yozin2015}). This would thus explain the UDG formation in high density environments. However, it fails to explain the presence of UDGs in lower densities or even in isolation. Therefore, UDGs have also been proposed to simply be dwarf-like galaxies that either live in high-spin halos that prevent objects becoming more condensed (\citealt{Amorisco2016}; \citealt{Rong2017}; \citealt{Amorisco2018}) or that are formed through internal processes, i.e. outflow-driven gas feedback that disperses the matter, creating a diffuse galaxy (e.g. \citealt{Chan2018}; \citealt{DiCintio2017}). Such dwarf-like scenarios predict the presence of UDGs in isolated environments, allowing for the presence of gas in such galaxies (e.g. \citealt{Jones2018}, \citealt{Leisman2017}; \citealt{Papastergis2017}). Therefore, in the normal-dwarf scenarios, UDGs would be created in dwarf-sized halos in low-density environments and would be later accreted into groups and clusters (e.g. \citealt{Safarzadeh2017}; \citealt{Carleton2018}), where physical processes such as ram--pressure stripping and starvation become more prominent. This would remove the gas reservoir and eventually quench the star formation in the galaxy. This idea is motivated by the finding that red spheroidal UDGs typically populate galaxy clusters, whereas bluer, irregularly shaped ones, are more commonly found in groups and in isolation (e.g. \citealt{Roman2017b}). Interestingly, there are both evidence and theoretical predictions supporting the competing scenarios, which suggest that the true nature of UDGs might not follow a unique formation pathway, but instead have a mixed-bag of origins. 

\begin{figure*}
\centering
\includegraphics[scale=0.55]{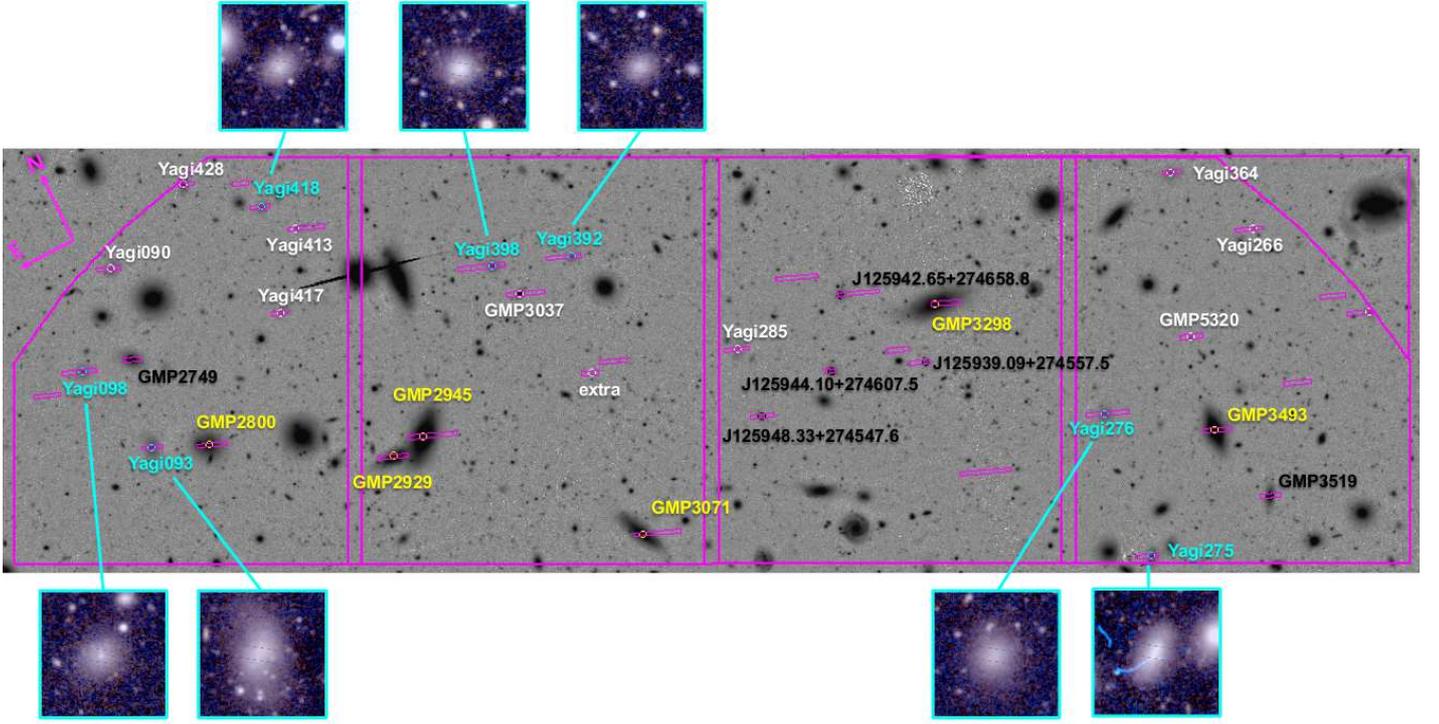}
\label{figure:1}
\vspace{-0.6cm}
\caption{\textbf{`Central' DEIMOS Coma mask}: Coma Cluster $V$-band Subaru/Suprime-Cam image \citep{Okabe2014} with the layout of the central DEIMOS mask (pink) shown. The mask covers roughly 16 $\times$ 4 sq. arcmin ($\sim$ 450 $\times$ 110\,kpc at the Coma distance) and is located at approximately 0.2$^{\circ}$ south of the centre of the Coma Cluster, with an angle of 160$^{\circ}$ (see Fig. 3 of Paper I). The cyan labelled galaxies correspond to the UDGs analysed in this work, whereas the yellow are high surface brightness galaxies and the black are dwarf galaxies, both used as control samples. The white labelled galaxies are the remaining galaxies that did not have enough S/N for a stellar population analysis but whose recession velocities are derived in Paper I. Unlabelled slits correspond to sky slits or alignment stars. The seven UDGs studied in this work are also shown in 30 $\times$ 30" thumbnails combining $V$ and $R_{\rm c}$ images.}
\end{figure*}

\begin{table*}
\centering
\label{table:1}                      
\begin{tabular}{l | l c c c c c c}   
\toprule      
\textbf{Galaxy}  &  \textbf{Type} & \textbf{R.A.} & \textbf{Dec.} & \textbf{Vr} &  \textbf{R} & \textbf{B--R} & \textbf{R$\mathrm{_e}$}\\  
                 &                           &    (J2000)    &   (J2000)     & (km\,s$^{-1}$)      &     (mag) &    (mag) & (kpc)    \\
\midrule
\midrule
Yagi093 (DF26) & UDG & 13:00:20.614 & +27:47:12.31 & 6611$\pm$\,137 & 18.9 & 0.96 & 3.49 \\
Yagi098        & UDG & 13:00:23.201 & +27:48:17.14 & 5980$\pm$\,82  & 19.6 & 0.96 & 2.30 \\
Yagi275        & UDG & 12:59:29.891 & +27:43:03.07 & 4847$\pm$\,149 & 19.2 & 0.92 & 2.93 \\
Yagi276 (DF28) & UDG & 12:59:30.463 & +27:44:50.40 & 7343$\pm$\,102 & 19.6 & 0.90 & 2.25 \\          
Yagi392        & UDG & 12:59:56.174 & +27:48:12.78 & 7748$\pm$\,161 & 20.7 & 0.97 & 1.46 \\          
Yagi398        & UDG & 13:00:00.414 & +27:48:19.68 & 4180$\pm$\,167 & 20.1 & 0.96 & 1.34 \\          
Yagi418        & UDG & 13:00:11.710 & +27:49:40.99 & 8335$\pm$\,187 & 20.4 & 0.93 & 1.57 \\          
\midrule                                                                                           
J125944+274607 & LLG (dE) & 12:59:44.105 & +27:46:07.57 & 6109$\pm$\,127 & 19.2 & - & 0.92 \\          
J125942+274658 & LLG (dE) & 12:59:42.650 & +27:46:59.44 & 5418$\pm$\,163 & 20.2 & - & 1.36 \\          
J125948+274547 & LLG (dE) & 12:59:48.372 & +27:45:48.21 & 8039$\pm$\,109 & 21.2 & - & 1.37 \\          
J125939+274557 & LLG (dE) & 12:59:39.096 & +27:45:57.53 & 7791$\pm$\,164 & 20.1 & - & 0.88 \\
GMP\,2749           & LLG (dE) & 13:00:20.482 & +27:48:17.03 & 5846$\pm$\,74  & 18.4 & - & 1.59 \\
GMP\,3519           & LLG (S)  & 12:59:22.944 & +27:43:24.48 & 4062$\pm$\,167 & 18.7 & - & 1.88 \\
\midrule                                                         
GMP\,2800 & HLG (dE0 )  & 13:00:17.553 & +27:47:03.94 & 7001$\pm$\,132 & 16.7 & - & 2.92 \\
GMP\,2923 & HLG (SBb )  & 13:00:08.054 & +27:46:24.08 & 8652$\pm$\,125 & 16.8 & - & 2.09 \\     
GMP\,2945 & HLG (Sa  )  & 13:00:06.288 & +27:46:32.88 & 6091$\pm$\,66  & 14.6 & - & 2.63 \\
GMP\,3071 & HLG (Sb  )  & 12:59:56.112 & +27:44:46.72 & 8810$\pm$\,99  & 16.2 & - & 1.39 \\
GMP\,3298 & HLG (S0/a)  & 12:59:37.828 & +27:46:36.62 & 5554$\pm$\,41  & 15.3 & - & 4.27 \\  
GMP\,3493 & HLG (Sa  )  & 12:59:24.931 & +27:44:19.86 & 6001$\pm$\,80  & 14.9 & - & 1.38 \\
\bottomrule
\end{tabular}
\vspace{0.1cm}
\caption{\textbf{Ultra-diffuse galaxy sample and other Coma cluster targets}. Summary of the observational and main properties of the UDGs and other galaxies studied in this work. For the 7 UDGs we use the ID from \citet{Yagi2016} and quote if they also have a Dragonfly name \citep{vanDokkum2015a}. For the other galaxies, we use their most common name as identified in the SDSS. The galaxy type is also shown, as quoted in either SIMBAD or NED, with their corresponding coordinates. The quoted recession velocities, $R$-band magnitudes, the $B-R$ colours and effective radii are as quoted in Paper I (from \citet{Yagi2016} for the UDGs and SDSS/SIMBAD/NED for the control galaxies). We separate between LLG and HLG at $R$=17.}
\end{table*}

There is, however, one key diagnostic to understand the origins of UDGs that has remained quite elusive to date -- the study of their stellar populations. By comparing such stellar populations to those expected from the diverse formation scenarios one should be able to differentiate between the possible origins. For example, under the `failed' galaxy scenario where the primordial halos were quenched at very early times, one would expect stellar populations with very old ages due to the very early quenching and high mean [Mg/Fe] ratios related to the fast formation timescales, e.g. \citet{Yozin2015}. In contrast, favouring a normal-dwarf galaxy interpretation, \citet{Rong2017} predicted from their cosmological simulations that UDGs formed late in halos of high spin, with extended star formation histories and a mean age of 7\,Gyr (some 2.5\,Gyr younger than the typical dwarf galaxy in their simulation). Similarly, analysing UDGs with high levels of feedback but normal spin parameters within the FIRE simulations, \citet{Chan2018} suggested that UDGs can have a range of quenching timescales, thus predicting a broad range of UDG ages, i.e. 2--12\,Gyr. 

Observationally, the majority of studies have addressed the issue of UDG stellar populations from a photometric point of view (e.g. \hypertarget{P+18}{\citealt{Pandya2018}}, hereafter \hyperlink{P+18}{P+18}; \citealt{Roman2017a}; \citealt{Trujillo2017}) or by studying their globular cluster properties (e.g. \citealt{BeasleyT2016}; \citealt{Peng2016}; \citealt{vanDokkum2017}; \citealt{Toloba2018}; \citealt{vanDokkum2018a}; \citealt{vanDokkum2018b}). Owing to the low surface brightness and extended size of UDGs, spectroscopy is very challenging and time-consuming, but crucial for revealing the formation histories of these objects. To date, only a couple of studies have attempted such a task. \citet{Kadowaki2017} obtained spectra of 4 Coma UDGs using the Large Binocular Telescope. However, due to the low signal-to-noise (S/N) of their spectra they had to stack them into one single spectrum, which was then visually compared to stellar population models. They concluded that their Coma UDGs were most compatible with being old and very metal-poor ([Fe/H]$<$ -1.5). By combining many fibres on the Apache Point Observatory telescope and a long integration time,  \hypertarget{G+18}{\citet{Gu2018}} (hereafter \hyperlink{G+18}{G+18}) were able to analyse individual spectra of 3 other Coma UDGs. From a spectral and photometric analysis, they concluded that all three galaxies were relatively old ($\sim$9 Gyr) and slightly metal-poor ([Fe/H] $\sim$ -1). Upon submission of this paper, a new spectroscopic study of three more UDGs in Coma (plus two in common to our work) has been presented by \hypertarget{RL+18}{\citet{Ruiz-Lara2018}} (hereafter \hyperlink{RL+18}{RL+18}), finding that the UDGs have intermediate to old ages ($\sim$7 Gyr) and low metallicites ([Z/H] $\sim$ -1).  

Here we extend the characterisation of high density environment UDGs by presenting the stellar populations of 7 UDGs that have been spectroscopically confirmed as Coma cluster members from new Keck/DEIMOS spectra. We combine this with literature data to examine the distributions of age, metallicity and $\alpha$--element overabundance for Coma cluster UDGs, and to see how they compare to the properties of other galaxies in Coma. With the additional information revealed by the star formation histories (SFHs) of our UDGs, we discuss the implications for their formation mechanisms and possible origins.  

Section 2 presents a summary of the new Keck/DEIMOS data, the sample, the observations and the data reduction, although the reader is encouraged to read the first manuscript of the series (\citet{Alabi2018}, Paper I hereafter) for a full description. Section 3 describes the stellar population analysis performed and we discuss the possible origins for the Coma UDGs in section 4. We assume a $\Lambda$CDM cosmology with $H_0$=70 km s$^{-1}$ Mpc$^{-1}$, $\Omega_m$=0.27 and $\Omega_\Lambda$=0.73 to allow for direct comparisons with the theoretical predictions and literature data, with an adopted distance for Coma of 100\,Mpc.

\begin{figure}
\centering
\includegraphics[scale=0.63]{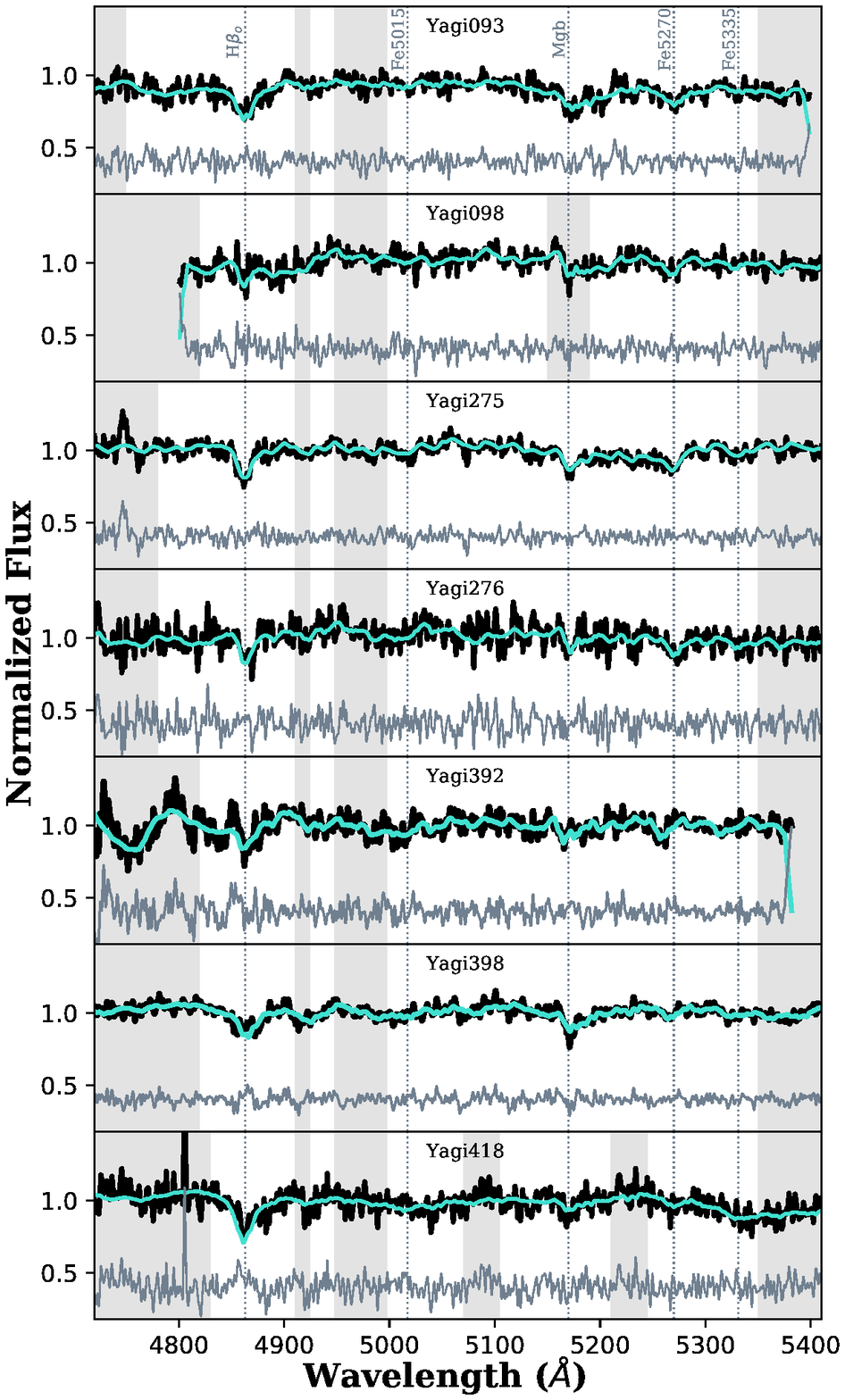}
\label{figure:2}
\vspace{-0.3cm}
\caption{\textbf{Coma UDG spectra}. Reduced spectra for the 7 UDGs presented in this work, highlighting the spectral range used for the stellar population analysis. The spectra and the full spectral fitting result (cyan line) have been normalised for illustration purposes. The grey shaded areas correspond to the masks used in the full spectral-fitting approach and the residuals from the fit are shown at the bottom of each panel in grey. The key line indices used in the index approach have been also labelled and marked by dotted lines.}
\end{figure}

\section{DATA}
\subsection{Sample selection} 
A detailed description of the selected candidates and DEIMOS mask production can be found in Paper I, but we summarise here the most relevant points of the sample selection. Candidate UDGs were selected from the catalog of \citet{Yagi2016}, based on deep Subaru/Suprime-Cam imaging of the Coma cluster \citep{Okabe2014}. Using criteria that included half-light radii of $\mathrm{R_e}>$\,0.7\,kpc (a factor of 2 smaller than Dragonfly UDGs of \citealt{vanDokkum2015a}), total magnitude -17\,$<\,\mathrm{M_R}\,<$\,-9, and a mean surface brightness within 1$\mathrm{R_e}$ of 24\, $<\,\mu\mathrm{_R}\,<$ 27\,mag\,arcsec$^{-2}$, they identified 854 Coma cluster candidate UDGs. In order to target as many UDGs as possible we focused on a 16 $\times$ 4 sq. arcmin region (i.e. the footprint of a DEIMOS mask) at approximately 0.2$^{\circ}$ ($\sim$340\,kpc) south of the centre of the Coma cluster (see Fig 3 of Paper I for reference). This mask, which is shown in Figure 1 and which we call `central' compared to the more external one from Paper I, contained a total of 50 slits - half of them targeting UDGs. The rest of slits were filled with high surface brightness galaxies of different types: high-luminosity galaxies (HLG; $R>$17) that are typically late-type ones, and low-luminosity galaxies (LLG; $R<$17) which are typically dwarf galaxies. 

\subsection{Observations and data reduction} 
Observations were carried out using the DEIMOS instrument on the Keck II telescope. Over 3 nights in 2017 (April 27 -29) we obtained a total of 29 individual exposures of 30 min each, giving a total on source exposure time of 14.5 hr for this central Coma mask, with seeing conditions of 0.6-0.8" and generally clear skies. We used the 600 lines\,mm$^{-1}$ grating centred at 6000\AA, which delivers a wavelength coverage spanning $\sim$4300 - 9600\AA  \,depending on the position of the slit. Each slit was opened 3" to match to the typical size of Coma UDGs. However, this configuration was too coarse to measure internal velocity dispersions, as it results in a FWHM resolution of $\sim$14\AA ~(or $\sigma$ $\sim$300 km\,s$^{-1}$ at the central wavelength). 

The raw data were reduced using the {\tt spec2D} pipeline \citep{Cooper2012}. As we are dealing with faint sources, we experimented with different object definition and sky-subtraction approaches but ultimately adopted the default procedure of the {\tt spec2D} pipeline as preferred. The output 1D spectra were sky-subtracted and wavelength calibrated. We also applied a relative flux calibration that corrects for the spectral shape using a set of standard stars that were also observed with the same setup. Paper I presents the analysis of the recession velocities of the galaxies in the mask, confirming them as Coma members. We thus used those published values to shift the spectra to the rest frame before performing any stellar population analysis. Because the instrumental dispersion already matched the stellar population models used in Section 3, there was no need to further broaden our spectra. 

Although the central mask contained 15 slits targeting candidate UDGs, only 7 of them had sufficient signal-to-noise (S/N\,$>$15; see \citealt{CidFernandes2014}) for a tentative stellar population analysis. In Table 1 we list the coordinates, recession velocities (as derived in Paper I), $R$-band magnitudes, colours, and effective radii from \citet{Yagi2016} for the 7 UDGs. Two of them are found in common with the \citet{vanDokkum2015a} Dragonfly catalog: Yagi093=DF26 and Yagi276=DF28. Interestingly, only one of the seven UDGs (Yagi093) shows some indications of being disrupted, which is in agreement with the general lack of tidal features in UDGs in the central parts of Coma (e.g. \citealt{Mowla2017}, and see similar results for the Perseus cluster; \citealt{Wittmann2017}). This is reinforced by the fact that Yagi093 is the largest and one of the furthest UDGs in our sample. Yagi398 is on the limit of the UDG size criteria defined by \citet{vanDokkum2015a} (with $\mathrm{R_e}$=1.3\,kpc) but it is within the limits from \citet{Yagi2016} and those in the theoretical simulations. Therefore we keep this galaxy in the UDG class, although we will check if our results in Section 4 are affected by such assumption. 

The additional galaxies included in the mask that have sufficient S/N are also described in Table 1. Four of them are newly confirmed Coma dwarf galaxies, one of them, J125942.65+274658.8 with properties that resemble more to UDGs than dwarfs. Like with Yagi398, we keep this object in the original classification but check for inconsistencies in the results. Two other filler objects (GMP\,2800 and GMP\,3298) are found in common with the sample of Coma red dwarf galaxies from \hypertarget{S+09}{\citet{Smith2009}} (hereafter  \hyperlink{S+09}{S+09}) used in Section 4, for which we find consistent age and metallicity values in the following section. 

In Figure 2 we show the spectra of the 7 UDGs in this work, corresponding to the best spectral range used for the stellar population analysis in Section 3, which is virtually free of sky residual and instrument flexures that were not properly corrected during the reduction process. This spectral range is sufficient to perform the stellar population analysis as it encompasses the most relevant features needed for it, which are highlighted by the dotted vertical lines. Note that both the spectra and the fit shown have been normalised for illustration purposes.  

\begin{figure}
\centering
\includegraphics[scale=0.4]{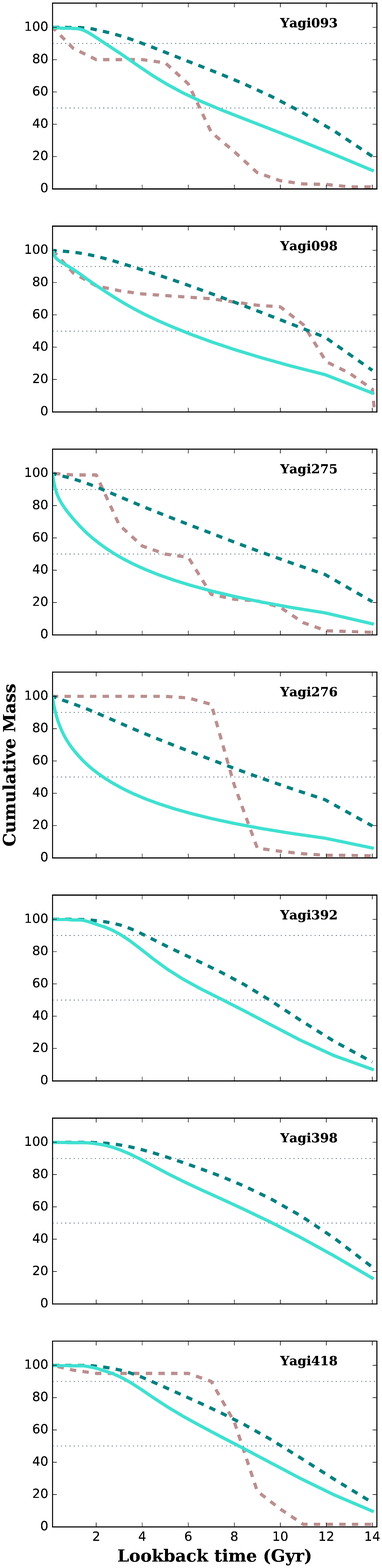}
\label{figure:3}
\vspace{-0.2cm}
\caption{\textbf{Star formation histories of Coma UDGs}. We show the amount of mass created over cosmic time (as look-back time) for the seven UDGs, representative of their SFHs. The cumulative mass  from {\tt STECKMAP} is represented by a cyan dashed line. For those UDGs with a good flux-calibration, the brown dashed line corresponds to the cumulative mass from {\tt STARLIGHT}. We also overplot the cumulative light used to derive the mean luminosity-weighted ages used for the UDGs. Each panel shows an individual UDG as labelled, with the horizontal dotted lines corresponding to 50$\%$ and 90$\%$ of mass.}
\end{figure}

\section{Stellar population analysis}
\subsection{Methodology}
For the stellar population analysis we employ the newest extension of the {\tt MILES} SSPs (Single-Stellar Population models; \citealt{Vazdekis2016}), which covers a wide range of ages, metallicities and initial mass functions. It has been shown that at low stellar masses and low velocity dispersions such as those of the objects studied here, the impact of a varying initial mass function is very mild (e.g. \citealt{Ferre-Mateu2013}), therefore we assume a universal Kroupa IMF to directly compare with literature results. In the following, we study the stellar population properties with a two-fold approach, using the best spectral range of our data (4700-5400\,\AA\,, see Fig. 2), which is virtually free from sky residuals, bad pixels and instrument flexure. 

We first perform a line-index analysis, using the age--sensitive indices H$\beta$ and H$\beta_{\rm o}$ compared to a set of metallicity indicators (Fe5015, Fe5270, Fe5335, Mgb, and also the composite indices <Fe>$^{\prime}$ , [MgFe50] and [MgFe]$^{\prime}$). This provides a set of luminosity-weighted ages, [Fe/H] metallicities, [Z/H] total metallicities and $\alpha$--abundances. However, because our spectral range only contains the $\alpha$ element Mg, we use the [Mg/Fe] abundance as the [$\alpha$/Fe] one (see Appendix for a full description on the line-indices used). We also use the full spectral fitting approach, which creates a combination of SSP model predictions that best matches each spectrum. This provides not only mean luminosity-weighted values but also mass-weighted ones and the SFHs of the galaxies over time, i.e. the amount of stellar mass/light that was created over cosmic time. 

Stellar ages and metallicities can also be obtained with the full spectral fitting approach. The advantage of this method is that it provides both mean luminosity- and mass-weighted values that are computed as the mean of the SFH of the galaxy. In general, the luminosity-weighted value should be similar to the one obtained from the line--index approach. For this exercise we use two different codes: one for the relative-flux calibrated spectra ({\tt STARLIGHT}; \citealt{CidFernandes2005}) and another applicable to non-flux calibrated spectra ({\tt STECKMAP}; \citealt{Ocvirk2006}). Despite providing similar results in terms of mean quantities, each code delivers the SFH in a different way, i.e. {\tt STARLIGHT} tends to produce SFHs that are more bursty-like whereas {\tt STECKMAP} normalises the result so it provides a more continuous SFH via regularisation. One can also interpret the SFHs by transforming them into cumulative fractions of mass and light, as shown in Figure 3. We can see that in general these galaxies built-up their mass and light in similar timescales, which means that their mean luminosity- and mass-weighted ages will be very similar (i.e. a difference of 1 or 2\,Gyr older for the mass-weighted values). However, note that Yagi098, Yagi275 and Yagi276 show virtually different cumulative trends, having luminosity-weighted age relatively younger than the rest, while their mass-weighted values are compatible with the rest of our UDGs. Nonetheless, such luminosity-weighted values are compatible with those derived with the line--index approach. Ideally, one would use the mass-weighted values as they better represent the true age of the stars. However, the values typically used in the literature are mostly luminosity-weighted, and therefore from here onwards we will use the  luminosity-weighted ages to allow for a fair comparison with the literature (Section 4.1). The figure also shows the cumulative mass from {\tt STARLIGHT} (dashed brown line) for comparison. Despite the differences between methods, all galaxies show extended formation histories that occur down to recent times. Such cumulative masses are compatible with the results from the SFHs of \hyperlink{RL+18}{RL+18}, although the data does not allow to discriminate between a bursty-like star formation, i.e. \textit{ala} {\tt STARLIGHT}, or a more continuous one, i.e. \textit{ala} {\tt STECKMAP}, as discussed by these authors. 

We can then estimate at what cosmic time did the galaxy form 50$\%$ and 90$\%$ of its stellar mass (t$_{50}$ and t$_{90}$, respectively) from their cumulative mass profiles. We  transform such stellar ages into the timescales the galaxy needs to build-up such stellar masses as $\Delta t_{50}$ = t$_{BigBang}$--t$_{50}$ and $\Delta t_{90}$ = t$_{50}$--t$_{90}$. Comparing such mass-weighted timescales can provide a sense of how fast/slow and how extended the star formation was. Additionally, we consider t$_{90}$ as a proxy for the `quenching time' of the galaxy, when star formation completely ceases. Note that we can not determine the reason why such quenching happens solely from the SFHs, therefore additional information is needed. 

Having a variety of measurements allows for flexibility in the results. This is, depending on the quality of the spectra one can choose which approach is more reliable. For example, if there is a noise spike near an absorption line relevant for the analysis, then the line index approach will not be reliable for that galaxy, while a poor flux calibration will not allow one to trust the derived SFH. A full detailed description of the stellar population analysis and some comparisons within the different methods can be found in the Appendix. After a meticulous inspection of each line-index used and spectral fit quality, we quote in Table 2 the most robust stellar population properties that will be used throughout the following sections. In order to be as consistent as possible, we use the {\tt STECKMAP} values (when possible) for the luminosity-weighted ages and total metallicities (as it uses the non-calibrated data, this method is less affected to possible issues resulting from calibrating MOS data). However, we use the results from the line--index approach for the [Fe/H] and $\alpha$--abundances. With these chosen stellar population values, we then derive the stellar mass of each UDG using their stellar population mass--to--light (M/L) ratio and the $R$--band total magnitude.   

\subsection{Results}
We find a varied range of stellar population properties, in particular for the ages and the abundances. The mean-luminosity weighted ages of our seven UDGs cover intermediate to old ages (4--8\,Gyr), with a mean age of 6.7\,$\pm$1.6\,Gyr. All the UDGs in this work have low total metallicities with a mean [Z/H]\,=\,-\,0.66\,$\pm$0.27 and a mean [Fe/H]\,=\,-\,0.87\,$\pm$0.79\,dex. The mean abundance ratio [Mg/Fe] \, is \,0.13\,$\pm$0.52, which is weighted down for the only UDG with under solar value (the rest all have super-solar values). Unfortunately, both [Fe/H] and [Mg/Fe] could only be measured for four out of the seven UDGs from the line indices. We additionally examined two other Coma UDGs (Yagi285/DF25 and Yagi413) in our sample but are not included in the analysis because their S/N was too low for a reliable stellar population analysis. However, to first order they were both consistent with having old ages and low metallicities.

From the {\tt STECKMAP} cumulative mass we can see that our UDGs took timescales of $\sim$2--5\,Gyr to build up half of their stellar mass (mean $\Delta t_{50}=$3.8$\pm$0.8\,Gyr). After that, all the UDGs maintain a steady formation rate with rather longer timescales of formation (mean $\Delta t_{90}=$6.5$\pm$0.3\,Gyr). If we were to use the {\tt STARLIGHT} results, we find slightly longer timescales to reach half of their stellar mass (mean $\Delta t_{50}=$5.0$\pm$0.9\,Gyr), which are mostly due to their delay of $\sim$2\,Gyr to start forming stars. Although the {\tt STARLIGHT} $\Delta t_{90}$ are shorter (4.4$\pm$1.6\,Gyr), the mean `quenching' age is the same for both methods, t$_{90}$=\,3.6\,Gyr. Note that as such ages and timescales do not have an associated error, we compute the standard error of the mean to account for the uncertainties in the measurements. We will further discuss the implications of such possible quenching timescales later in Section 4.2.

There is, however, a caveat with the use of line indices (and the study of integrated light, in general) to retrieve mean luminosity ages. The existence of old, metal poor ([Z/H]$\lesssim$-1\,dex or [Fe/H]$\lesssim$-1.5\,dex) bright stars, such as blue horizontal branch (BHB) stars, can mimic the signatures of young stars by boosting the equivalent width of the Balmer lines. In these cases, it is hard to distinguish between a truly young stellar population and an old one that hosts BHB stars. Such an effect has been reported in globular clusters and for some dwarf galaxies in groups (e.g. \citealt{Monaco2003};  \citealt{Schiavon2004}); \citealt{Ocvirk2010}; \citealt{Deason2015}; \citealt{Conroy2018}) but it is still unclear for more massive systems such as the ones studied here. Hence, if BHB stars were present in these galaxies, it is conceivable that they could actually be older than what we measure. We briefly investigate if our stellar ages and SFHs could be affected by this issue.

First, we can see from our spectra that none of the UDGs have a remarkable H$\beta$ line that could indicate a very young age, not even for the two youngest UDGs in our sample. Second, the younger UDGs are the ones we would be most worried about, however they all show higher metallicities, which are known to be free of such an effect (e.g. \citealt{Ocvirk2010}). Only two of our UDGs could potentially be affected, as they have metallicities lower than or similar to --1\,dex (Yagi398 and Yagi418). However, they are already the oldest UDGs in our sample, and thus such effect would be milder. At most, the ages for these two UDGs would be considered lower limits. In fact, if any object could be affected it would be the LLGs in our sample. With metallicities typically below $-$1\,dex, the majority present intermediate age populations and strong H$\beta$ features. Third, some line index diagnostics can be used to discriminate this effect (e.g. \citealt{Schiavon2004}). Unfortunately, our spectral range does not cover such diagnostic tools. However, it has been shown that the use of some full-spectral-fitting techniques, in particular {\tt STECKMAP}, can overcome this issue. It has been proven that the presence of BHB stars would only contribute to an inferred recent burst of star formation at $\sim$100\,Myr at levels of less than 10$\%$ in light (\citealt{Ocvirk2010}). The derived SFHs of our objects are all extended, with the youngest episodes of formation occurring $\sim$1\,Gyr ago at most. Considering all these fact together, we caution the reader about the possible effects but proceed the analysis with the reported stellar population values.

\begin{table*}
\centering
\label{table:2}                      
\begin{tabular}{p{0.85in} | p{0.1in} p{0.1in} c c c c c c c c c} 
\toprule       
Galaxy            &              & S/N   & \textbf{Age} & \textbf{[Fe/H]}      & \textbf{[Z/H]}           & \textbf{[$\alpha$/Fe]} & \textbf{t$_{50}$} & \textbf{t$_{90}$} &\textbf{$\Delta$t$_{50}$} & \textbf{$\Delta$t$_{90}$} &\textbf{M$_{\ast}$}\\  
            &                  &       & (Gyr)        &  (dex)               &   (dex)                  &      (dex)             & (Gyr)             & (Gyr)             &(Gyr)             & (Gyr)             &(M$_{\odot}$) \\
\midrule                                                                                      
\midrule                                                                                      
Yagi093        & (3) & 23 &  7.9$\pm\,$1.8 & $-$1.48$\pm\,$0.82 & $-$0.56$\pm\,$0.16 &  0.64$\pm\,$0.25   & 10.8 & 4.1  &  3.2 &  6.7 & 3.1E+08\\  
Yagi098        & (3) & 19 &  6.7$\pm\,$2.6 &    --              & $-$0.72$\pm\,$0.18 &  --                & 11.2 & 3.5  &  2.8 &  7.7 & 1.1E+08\\  
Yagi275        & (3) & 25 &  4.6$\pm\,$2.6 & $-$0.06$\pm\,$0.51 & $-$0.37$\pm\,$0.19 & $-$0.42$\pm\,$0.65 &  9.3 & 2.2  &  4.7 &  7.1 & 9.4E+07\\  
Yagi276        & (3) & 18 &  4.2$\pm\,$2.3 &    --              & $-$0.38$\pm\,$0.71 &  --                &  9.1 & 2.0  &  4.9 &  7.1 & 1.4E+08\\  
Yagi392        & (3) & 15 &  7.4$\pm\,$2.1 &    --              & $-$0.58$\pm\,$0.28 &  --                &  9.7 & 4.1  &  4.3 &  5.6 & 9.1E+07\\ 
Yagi398        & (3) & 21 &  8.3$\pm\,$3.1 & $-$0.48$\pm\,$0.87 & $-$0.92$\pm\,$0.38 &  0.06$\pm\,$0.68   & 11.2 & 5.2  &  2.8 &  6.0 & 3.6E+07\\   
Yagi418        & (3) & 18 &  7.9$\pm\,$2.0 & $-$1.48$\pm\,$0.96 & $-$1.10$\pm\,$0.95 &  0.27$\pm\,$0.53   & 10.0 & 4.4  &  4.0 &  5.6 & 1.1E+08\\  
\midrule                                                                                                                      
\rowcolor{BlueGreen}
\textbf{Mean UDGs}&&      &  6.7$\pm\,$1.6  & $-$0.87$\pm\,$0.79 & $-$0.66$\pm\,$0.27 &  0.13$\pm\,$0.52   & 10.2$\pm\,$0.3 &  3.6$\pm\,$0.4 &  3.8$\pm\,$0.8 &  6.5$\pm\,$0.3 & 1.3E+08\\
\midrule
\midrule
J125944+274607 & (3) & 18 &  8.7$\pm\,$4.0 & $-$0.90$\pm\,$0.97 & $-$1.01$\pm\,$0.10 & $-$0.06$\pm\,$0.40 & 11.0 &  8.3 &  3.0 &  2.7 & 2.1E+08 \\ 
J125942+274658 & (1) & 15 &  6.0$\pm\,$2.0 &   --               & $-$1.38$\pm\,$0.50 &  --                &  --  &   -- &  --  &  --  & 1.7E+08 \\ 
J125948+274547 & (2) & 15 &  7.6$\pm\,$2.0 &   --               & $-$0.89$\pm\,$0.50 &  --                & 10.0 &  1.3 &  4.0 &  8.7 & 5.1E+07 \\ 
J125939+274557 & (3) & 18 & 10.7$\pm\,$1.1 &   --               & $-$1.26$\pm\,$0.18 &  --                &  9.3 &  7.1 &  4.7 &  2.2 & 2.4E+08 \\ 
GMP\,2749      & (3) & 30 &  6.8$\pm\,$2.7 & $-$0.76$\pm\,$0.49 & $-$0.87$\pm\,$0.09 &  0.08$\pm\,$0.85   &  8.3 &  7.4 &  5.7 &  0.9 & 2.6E+08 \\ 
GMP\,3519      & (3) & 25 &  7.9$\pm\,$1.0 & $-$0.86$\pm\,$0.16 & $-$0.89$\pm\,$0.15 & $-$0.14$\pm\,$0.17 &  7.5 &  3.9 &  6.5 &  3.6 & 6.5E+08 \\ 
\midrule                                                                                                               
\rowcolor{Gray}
\textbf{Mean LLGs}&&      & 7.9$\pm\,$1.8 & $-$0.84$\pm\,$0.23 & $-$1.05$\pm\,$0.28 & $-$0.04$\pm\,$0.47 &  9.2$\pm\,$0.5 &  5.6$\pm\,$1.1 &  4.8$\pm\,$0.5 &  3.6$\pm\,$1.1 & 2.7E+08 \\
\midrule                                                                                              
\midrule                                                                                              
GMP\,2800      & (3) & 40 & 9.76$\pm\,$0.5 & $-$0.34$\pm\,$0.21 & $-$0.34$\pm\,$0.20 & $-$0.14$\pm\,$0.43 &  4.7 &  4.0 &  9.3 &  0.7 & 1.0E+10 \\           
GMP\,2923      & (3) & 38 & 6.72$\pm\,$1.3 & $-$0.45$\pm\,$0.38 & $-$0.45$\pm\,$0.27 & $-$0.06$\pm\,$0.70 & 10.0 &  1.2 &  4.0 &  8.8 & 9.8E+08 \\           
GMP\,2945      & (3) & 40 & 9.25$\pm\,$4.0 & $-$0.27$\pm\,$0.30 & $-$0.27$\pm\,$0.28 &  0.08$\pm\,$0.35   & 13.0 & 12.0 &  1.0 &  1.0 & 1.9E+10 \\           
GMP\,3071      & (2) & 50 & 8.65$\pm\,$0.3 & $-$0.63$\pm\,$0.24 & $-$0.63$\pm\,$0.44 &  0.47$\pm\,$0.43   & 13.0 &  1.4 &  1.0 & 11.6 & 3.7E+09 \\           
GMP\,3298      & (2) & 55 & 9.45$\pm\,$2.4 & $-$0.33$\pm\,$0.12 & $-$0.33$\pm\,$0.11 &  0.19$\pm\,$0.14   & 11.0 &  3.5 &  3.0 &  7.5 & 5.0E+09 \\           
GMP\,3493      & (3) & 53 & 9.40$\pm\,$2.4 & $-$0.27$\pm\,$0.33 & $-$0.27$\pm\,$0.22 &  0.27$\pm\,$0.39   &  4.7 &  4.0 &  9.3 &  0.7 & 4.0E+09 \\        
\midrule    
\rowcolor{Goldenrod}                                                                                                      
\textbf{Mean HLGs}&&      & 8.9$\pm\,$1.6 & $-$0.38$\pm\,$0.26  & $-$0.31$\pm\,$0.21 & 0.14$\pm\,$0.40    & 9.4$\pm\,$1.5 &  4.3$\pm\,$1.6 & 4.6$\pm\,$1.5  &  5.0$\pm\,$1.6 & 7.1E+09\\
\bottomrule
\end{tabular}
\vspace{0.1cm}
\caption{\textbf{Stellar Population Properties of Coma Galaxies.} The table presents the most robust results for the stellar population properties, as discussed in the Appendix. It shows the individual values for our DEIMOS mask UDGs, LLGs (low-luminosity galaxies) and HLGs (high-luminosity galaxies) but also the mean values for each type. First column specifies the method used to derive the luminosity-weighted ages and total metallicities (1=line--indices; 2={\tt STARLIGHT}; 3={\tt STECKMAP}), whereas [Fe/H] and [Mg/Fe] values are always derived from the line indices. The table also quotes the derived S/N within the spectral coverage used for the full-spectral-fitting. It presents the ages when the galaxy achieved 50$\%$ and 90$\%$ of its stellar mass (t$_{50}$ and t$_{90}$, as lookback ages) and the timescales to achieve such masses ($\Delta t_{50}$ = ( t$_{BigBang}$--t$_{50}$) and $\Delta t_{90}$ = (t$_{50}$--t$_{90}$), as increments in time). Therefore these 4 columns are mass-weighted values. Lastly, it presents the stellar mass calculated with the stellar population M/L ratios. }
\end{table*}

\begin{figure*}
\centering
\includegraphics[width=1.0\textwidth]{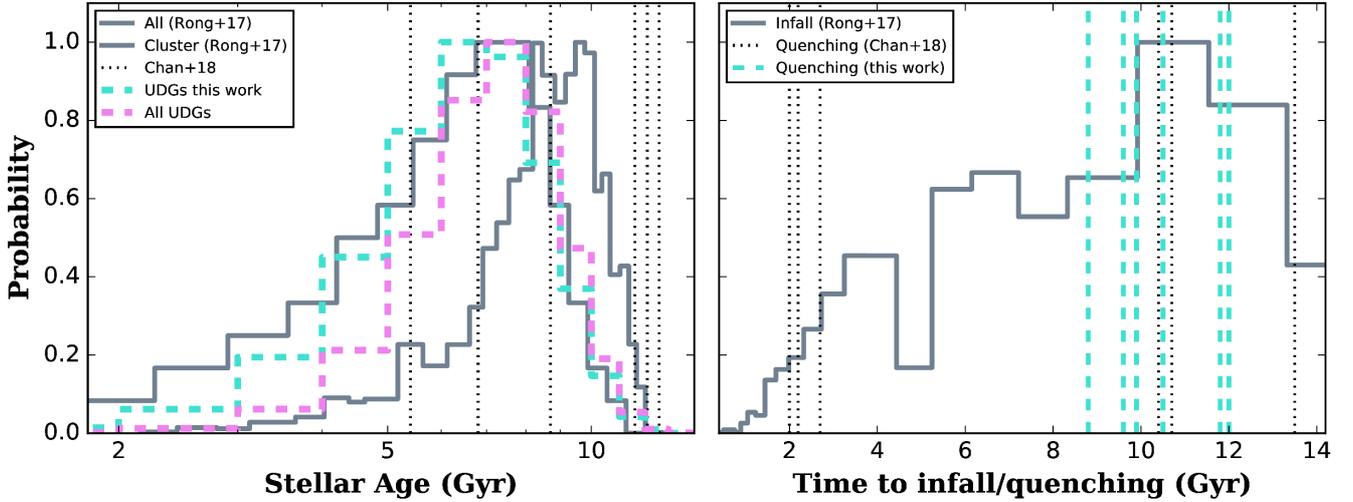}
\label{figure:4}
\vspace{-0.4cm}
\caption{\textbf{Theoretical predictions for UDGs}. \textit{Left panel:} The theoretical age distribution for UDGs from the simulations of \citet{Rong2017}, as expected if UDGs are dwarf-like galaxies. The dark grey histogram represents the distribution for all the simulated UDGs (cluster+field) whereas the light grey one corresponds to only cluster UDGs. The dotted black lines show the ages of the simulated UDGs in \citet{Chan2018}. Super-imposed we show the observed distribution of ages for our 7 Coma UDGs (dashed cyan histogram) and the distribution if we include other literature UDGs from G+18 and P+18 (dashed purple histogram). \textit{Right panel:} Predictions for the infall time of the \citealt{Rong2017} UDGs (grey histogram), showing that their simulated UDGs tend to have late infalling times. Overlaid are the quenching times for UDGs from the \citealt{Chan2018} simulations and the t$_{90}$ for our UDGs. Even if there is a $\sim$1.5-2\,Gyr delay between the time of infall and the galaxy quenching (e.g. \citealt{Muzzin2008}; \citealt{Haines2015}), all our UDGs are compatible with a late infall into the cluster environment.}
\end{figure*}

\begin{figure*}
\centering
\includegraphics[scale=0.60]{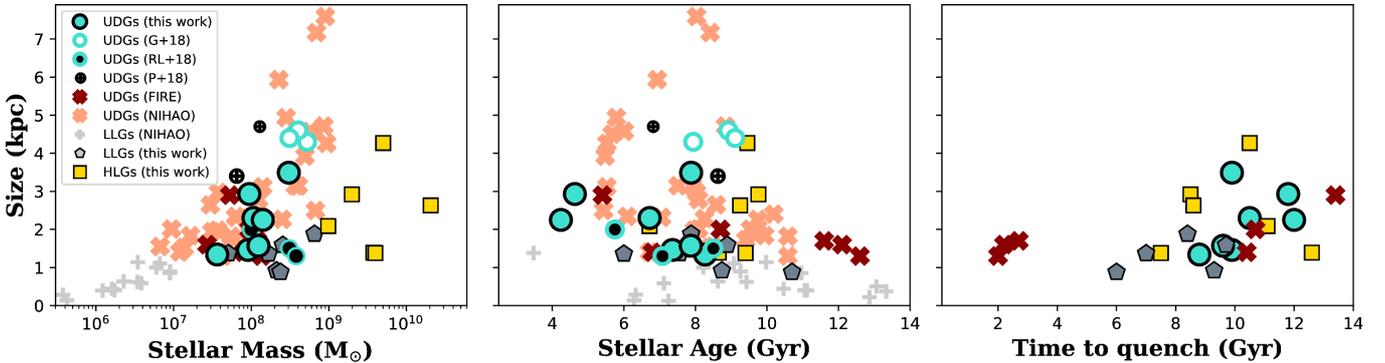}
\label{figure:5}
\vspace{-0.4cm}
\caption{\textbf{Sizes of simulated and observed UDGs}. \textit{Left panel} shows the mass--size relation for the simulated UDGs in both FIRE (red crosses; \citealt{Chan2018}) and NIHAO (pink crosses; \citealt{DiCintio2017}) simulations. NIHAO simulations for normal dwarf galaxies are also included (grey crosses, private communication). We compare such simulations to the observed galaxies in Coma from our mask: UDGs (cyan circles), dwarf galaxies (grey pentagons) and HLG galaxies (yellow squares). We also include other literature UDGs (open cyan circles for G+18; black-cyan circles for RL+18, crossed open circles for P+18 and with DGSAT\,I as a smaller symbol to show it is a field UDG rather than a cluster one). \textit{The middle panel} shows the age--size relation presenting a trend where younger UDGs have mildly more extended sizes, which could be related to their later infall into the cluster. \textit{The right panel} emphasises such trend from the age by showing that galaxies with longer quenching timescales are larger.} 
\end{figure*}

\begin{figure}
\centering
\includegraphics[scale=0.75]{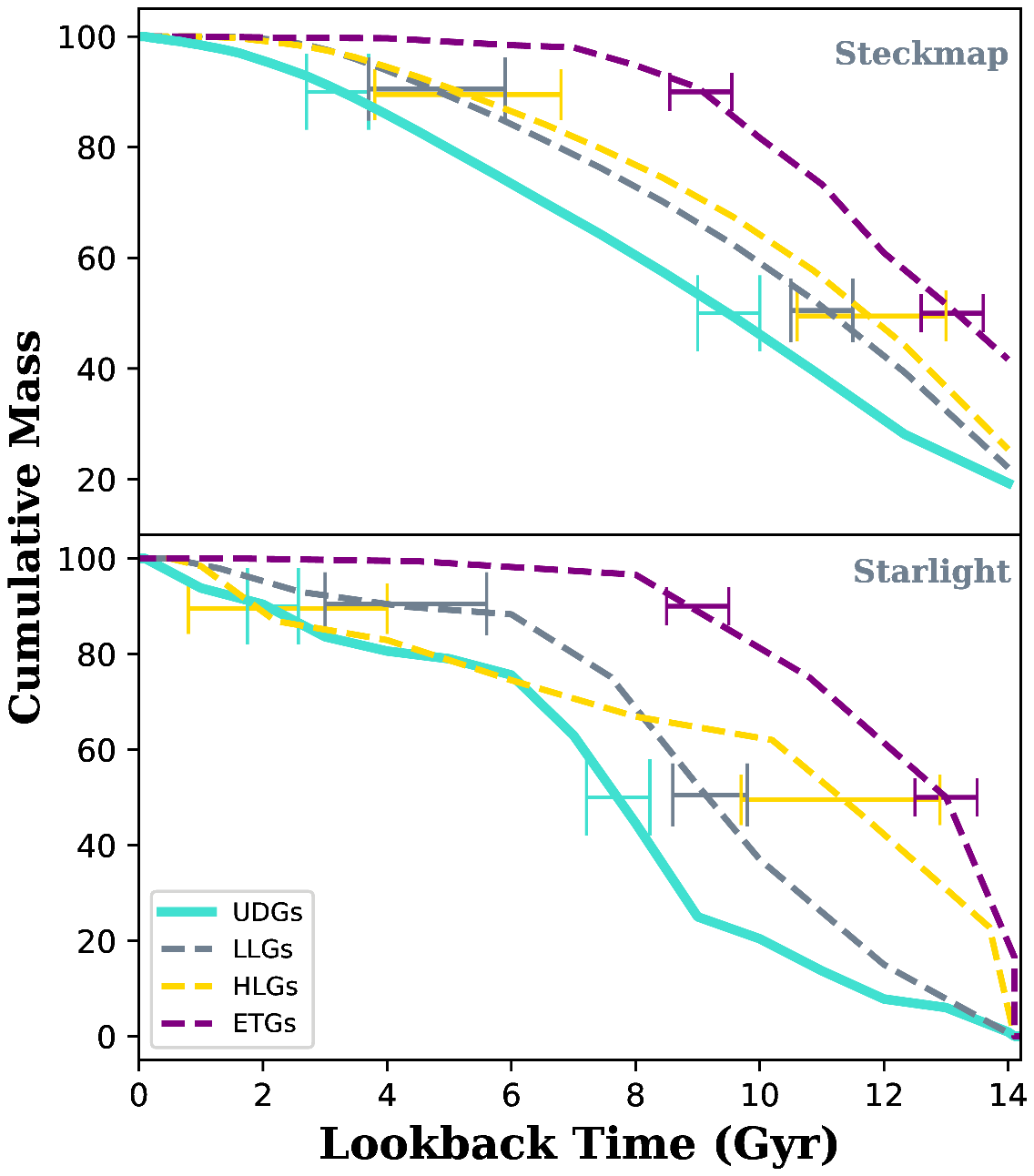}
\label{figure:6}
\vspace{-0.2cm}
\caption{\textbf{Mass assembly of Coma Galaxies by type}. The averaged SFHs for the different galaxy types in Coma are translated into the cumulative stellar mass, to show the different formation timescales the galaxies undergo. Top panel corresponds to the SFHs measured with {\tt STECKMAP} whereas the bottom panel shows the same SFHs obtained with {\tt STARLIGHT}. Despite the differences between both methods (see Section 3.1), the different formation timescales of each galaxy type follow similar trends. Our UDGs (cyan continuous line) present extended and steady SFHs with late formation epochs, not quenching until $\sim$2--3\,Gyr ago. The LLGs in the mask (grey dashed line) seem to follow the UDGs timescales closely, albeit they quenched $\sim$4--5\,Gyr ago. This $\sim$2\,Gyr delay of the UDGs is compatible with the theoretical model of \citet{Rong2017}. In contrast, both massive ETGs and normal HLG galaxies show early, steep formation epochs, with ETGs quenching at very early times and HLGs at later times compatible with the more extended SFHs seen for low-mass galaxies. Error bars correspond to the standard error on the mean computed using all the galaxies in each type to account for uncertainties in the ages and timescales of formation. }
\end{figure}

\section{Discussion}
We have presented new spectroscopic data for seven UDGs in Coma, the largest spectroscopic study of the stellar populations of UDGs to date. They are confirmed members of Coma, at a mean $z \sim$0.0223. They have rather red colours $\langle\,B-R\rangle\sim$0.94, an average size of $\langle\,\mathrm{R_{e}}\rangle\sim$\,2.3\,kpc and a stellar mass of $\langle\,M_{*}\rangle\sim$1.3$\times$10${^8}$\,M$_{\odot}$. As given in Table 2, we find a range of stellar population properties for our Coma UDGs, with a mean luminosity-weighted age of 6.7\,$\pm$\,1.6\,Gyr, total metallicity of [Z/H]\,=\,--\,0.66\,$\pm$0.27 and $\alpha$--abundance ratio of [Mg/Fe] \,=\,0.13\,$\pm$0.52. They all present extended SFHs, taking about 11\,Gyr to stop forming stars. 

As our aim is to reveal the origin(s) of these galaxies based on their stellar populations we combine our new data with other UDGs from the literature. We include spectroscopic data for the three Coma UDGs from \hyperlink{G+18}{G+18} (i.e. DF44, DF17 and DF07) and spectrophotometric data for two UDGs from \hyperlink{P+18}{P+18} (i.e. the Virgo member VCC\,1287 and the field UDG called DGSAT\,I). During this paper submission, another study with additional UDG spectroscopic data in a similar region of Coma was published (\hyperlink{RL+18}{RL+18}). We have two UDGs in common (Yagi093/DF26 and Yagi418), which allows for direct comparison with our results. \hyperlink{RL+18}{RL+18} reported one additional \citet{Yagi2016} UDG (Yagi090) and two newly discovered ones (OGS1 and OGS2). We caution the reader that while the later two UDGs are slightly outside the UDG definition of both \citet{Yagi2016} and \citet{vanDokkum2015a}, we include them in the comparison due to the variable nature UDGs are showing to have. We also note that \hyperlink{RL+18}{RL+18} UDGs share a similar region in Coma to ours, whereas \hyperlink{G+18}{G+18} UDGs are further out from the Coma centre. Both DF07 and DF17 are also part of the spectroscopic sample of \citet{Kadowaki2017}, with four more Coma UDGs located at even larger cluster-centric radii (DF03, DF08, DF30 and DF40). We do not use their results in the following figures because their stellar populations were estimated by visual comparison, but we consider them for the discussion statements. Finally, It is worth emphasising that neither \hyperlink{G+18}{G+18} nor \hyperlink{P+18}{P+18} derived [Mg/Fe] ratios but  assumed solar values. 

Besides including the results from the LLG and HLG galaxies included in our DEIMOS mask, we also use a sample of Coma galaxies from \hyperlink{S+09}{S+09} that occupy the same area as our UDGs. They were initially selected to represent a red population of dwarf galaxies but they covered a range in luminosities. We therefore apply a similar cut to our LLG and HLG mask galaxies at $R$=17 to separate between low-luminosity dwarf galaxies and more luminous ones. Finally, to extend to more massive objects inhabiting the surrounding areas in the Coma cluster core, we also include a sample of Coma galaxies from \hypertarget{AFM+14}{\citet{Ferre-Mateu2014}} (\hyperlink{AFM+14}{AFM+14}  hereafter). They have been separated morphologically in \citealt{Sanchez-Blazquez2006a} between ETGs and spirals (which we lable as HLGs).  

\subsection{Comparison with other UDG literature}
We now discuss how our results compare to the scarce literature values published so far, and how such results fit into the theoretical expectations, with the final goal to reveal the origin (or origins) of our Coma UDGs. Our stellar ages are consistent with the recent studies of \hyperlink{RL+18}{RL+18} and \hyperlink{G+18}{G+18}, with luminosity-weighted ages ranging mostly from 6--10\,Gyr, low metallicities and super solar alpha enhancements. A way to test the robustness of our results is by comparing them with those in \hyperlink{RL+18}{RL+18} for the UDGs in common. For DF26 (Yagi093 in this work) we obtain t\,=\,7.9$\pm$1.8\,Gyr,  [Z/H]\,=\,-0.56$\pm$0.18 and [Mg/Fe]\,=\,0.64$\pm$0.25 whereas they measure t\,=\,6.8$\pm$1.2\,Gyr, [Z/H]\,=\,-0.78$\pm$0.08 and [Mg/Fe]\,=\,0.25. For Yagi418, we find  t\,=\,7.9$\pm$2.0\,Gyr,  [Z/H]\,=\,-1.10$\pm$0.95 and [Mg/Fe]\,=\,0.27$\pm$0.52 as opposed to their t\,=\,8.1$\pm$1.1\,Gyr,  [Z/H]\,=\,-1.25$\pm$0.05 and [Mg/Fe]\,=\,0.60, showing a fairly good agreement. They also show similar formation ages t$_{50}$ and t$_{90}$, with 11 \textit{vs} 12\,Gyr and 4.0 \textit{vs} 6\,Gyr for Yagi093, and 10  \textit{vs} 10\,Gyr and 4  \textit{vs} 4\,Gyr for Yagi418, respectively).  In fact, from Panel b in figure 13 of \hyperlink{RL+18}{RL+18} it is seen that all but one of their UDGs have cumulative masses with similar trends to those found here. The mean timescales $\Delta$t$_{50}$ and $\Delta$t$_{90}$ for our UDGs are $\sim$4 and 6\,Gyr, whereas their mean times are $\sim$3 and 5\,Gyr, respectively.

Similar ages but somewhat lower metallicities have been also reported for another cluster UDG in Virgo, VCC\,1287 (with a lower limit age of 8.6\,Gyr and and lower limit [Z/H]\,=\,-1.55; \hyperlink{P+18}{P+18}). Similarly low metallicities have also been found to be the best match for the 4 UDGs from \citet{Kadowaki2017}, although they only report their stacked spectra to be mostly compatible with a very old SSP. On the contrary, a higher metallicity of [Z/H]\,=\,-0.63$^{+0.35}_{-0.62}$ has been reported for a field UDG, DGSAT\,I (\hyperlink{P+18}{P+18}), with a younger age of 6.81$^{+4.08}_{-3.02}$\,Gyr that indicates a somewhat extended SFH, which is very similar to our UDG results.

\subsection{Comparison with theoretical predictions}
We find that UDGs (including the literature ones) show a range of stellar ages regardless of the methodology used, ranging from 4 to 9\,Gyr. We now investigate at how this age distribution compares with the theoretical predictions. Figure 4 shows the observed age distribution for our sample of UDGs alone (cyan dashed histogram) and also if we include the literature ones (purple dashed histogram). The simulated FIRE UDGs of \cite{Chan2018} have their ages marked as vertical dotted lines. The continuous histograms correspond to the theoretical prediction models from \citet{Rong2017}, both for the entire UDG population (dark grey, skewed towards low density environments) and only for the cluster environment (light grey). The figure shows an age distribution for our UDGs peaking at an age of $\sim$7--8\,Gyr that is strikingly similar to the one of the simulated field UDGs, although the distribution of cluster UDGs in the simulations are slightly older ($\sim$9\,Gyr). This is also compatible with the stellar population results from the semi--analytical model of \citet{Carleton2018}. Our UDGs reveal a range of ages that correspond to a redshift of formation of $z\lesssim$\,1, which would exclude any of the scenarios that expect very old ages (i.e. the `failed' galaxies scenario), where formation redshifts of at least $\gtrsim$2 are expected. 

As the mean luminosity-weighted ages can be biased towards younger bursts of star formation, we also look at the predictions from the quenching times due to the infall into the cluster environment. The right panel of Figure 4 shows the predicted distribution of infall time into the cluster from \citet{Rong2017}, showing typically very long timescales for their simulated UDGs (up to $\sim$\,10\,Gyr). We also overlay the times used in the FIRE simulations to quench theirUDGs. Note that this quenched time does not have to directly represent the infall time, as quenching might occur by internal processes before infall (e.g. as described by \citealt{DiCintio2017} to explain field UDGs) or be caused by external processes during the infall (see also e.g. \citealt{Safarzadeh2017}). In fact, in the FIRE simulations the gas removal mechanism is not specified but they assume that the hot gas reservoir in the feedback-expanded dwarfs will be removed by ram pressure stripping, while suffering strangulation or a feedback episode as they fall into the clusters. This will quench star formation and turn the UDG into a redder galaxy. Therefore, they predict a very wide range of quenching times, some happening at early times ($\sim$\,2\,Gyr after the Big Bang, which would indicate a `primordial-like' type) and some late ones (after $\sim$\,11\,Gyr, indicative of a `late-dwarf' type). In their simulations, more massive UDGs quench at earlier times than less massive ones. Furthermore, note that an infall into the cluster does not imply an instantaneous quenching but galaxies take, on average $\sim$1.5--2\,Gyr to quench as they suffer the above depicted interactions (e.g. \citealt{Muzzin2008};  \citealt{Haines2015}). Therefore the vertical lines in the right panel of Figure 4 should be shifted by that amount of time towards the left-side to better represent the possible `infall' time. This is the same case for our UDGs, which present substantial star formation rates down to recent times. What is considered as such `quenching time' for our UDGs, which occurred between only 2 and 5\,Gyr ago, is shown with vertical dashed cyan lines. The figure shows that all our UDGs are compatible with the distribution of \citet{Rong2017} with late infalls and to the late quenched UDGs of \citet{Chan2018}, even if we corrected for the delay between infall and quenching. This is also in agreement with the infall times reported in Paper I from their velocity phase-space analysis, where we show that these 7 UDGs are mostly late infalls.

We now compare our observed UDGs with simulated ones from both FIRE \citep{Chan2018} and NIHAO (\citealt{Wang2015}; \citealt{DiCintio2017}) simulations, in terms of their size. We can directly compare to the FIRE simulations as they force the galaxies to quench at different times due to cluster infall, having two types of simulated UDGs. One type of UDG quenches at earlier epochs and thus has old ages, which would be more indicative of the `primordial-like' origin. The other type of UDGs, freely evolves in low-density environments, quenching at later times, thus having younger ages. One expectation from this scenario is that old, early-quenched UDGs would have smaller sizes than their younger, late-quenched counterparts. Those in NIHAO are solely field galaxies that evolve through internal processes without suffering cluster infall. However, if our assumption of such a late infall onto the Coma cluster is correct, the results should not differ extremely from the NIHAO results. The left panel of Figure 5 shows the stellar mass--size relation for both simulations and observations, with the well-established trend of more massive galaxies being larger. Our Coma UDGs (filled cyan circles) are all compatible with both FIRE and NIHAO simulations, whereas larger UDGs are instead only reproduced by the largest NIHAO UDGs. This suggests that such galaxies could have had more time to evolve in low-density environments and have been accreted onto the cluster more recently. They correspond to the UDGs in the outer parts of Coma (open cyan circles) and the field UDG DGSAT\,I, further supporting this assumption. One of our UDGS, Yagi093 presents somewhat larger sizes to the rest of our Coma UDGs, being the only one with some hints of tidal features. It could therefore be more similar to these outer Coma UDGs than to the more central ones in our sample. Also note that the two newly reported UDGs of \hyperlink{RL+18}{RL+18} (OGS1 and OGS2) are more compatible with being low-luminosity objects than UDGs. The middle panel clearly differentiates the two types of UDGs in the FIRE simulations (early \textit{vs} late quenched), showing a trend where older galaxies are typically smaller, which is emphasised by the NIHAO simulated UDGs. This is further highlighted in the rightmost panel of Figure 5, which shows the relation of the galaxy size with the quenching timescales (how long did the galaxy need to halt its star formation). Those that quenched faster (i.e. took only a few Gyr) have small sizes whereas, as the quenching time becomes longer ($\gtrsim$10\,Gyr), UDGs become more extended.

We remind the reader that we are here comparing the observed data with simulations, which are mostly based on a `dwarf-like' scenario. In particular, we are comparing to simulations where the galaxies are quenched due to their infall into the cluster, although our data do not allow us to determine the cause of the UDG quenching. However, after comparing our results with such simulations and including the results from Paper I and other stellar population analysis of nearby UDGs, we believe that the assumption of them being quenched due to a late infall is the most plausible one.

Of special relevance to such a late infall scenario are the derived SFHs of the UDGs, the main new contribution of this work. We have found that these Coma UDGs present sustained rates of star formation down to very recent times, whereas if they had fallen into the cluster at earlier epochs they would have stopped forming stars a long time ago. Figure 6 shows the averaged SFH (using both methods) for our UDGs compared to the averaged SFHs of our Coma mask LLGs galaxies, HLGs and massive ETGs from  \hyperlink{AFM+14}{AFM+14}. This figure emphasises the different formation histories the diverse type of galaxies have undergone. Both massive ETGs and HLGs start building up their stellar masses at the earliest epochs, with very high star formation rates that allow them to reach t$_{50}$ in less than 1 and 2\,Gyr, respectively. In contrast, both our control LLGs and UDGs start to build up their stellar mass at slower rates. UDGs and LLGs require around $\sim$6 and 5\,Gyr, respectively, to build up half of their mass. After they have created half of their mass, LLGs required an additional $\sim$5\,Gyr to reach 90$\%$ of their mass. Our UDGs require instead about $\sim$6 more Gyr after reaching half their stellar mass, quenching only $\sim$2\,Gyr ago. This difference of $\sim$2\,Gyr between the `quenching times' of UDGs and our LLGs is equivalent to the difference in the infall time between UDGs and the dwarf galaxies of \citet{Rong2017}, which further reinforces the late infall scenario proposed above. The question that arises is whether the location within the cluster of these objects is compatible with such a recent infall assumption, which we discuss next.

\subsection{Stellar populations dependence on cluster-centric radius}
We next study if there is any dependence of the properties with cluster-centric radius that could provide a hint about the infall/quenching times. Figure 7 shows the mean luminosity-weighted age and total metallicity with projected distance for our sample of Coma UDGs, \hyperlink{G+18}{G+18} and \hyperlink{RL+18}{RL+18} UDGs, our mask control galaxies, and the \hyperlink{S+09}{S+09} and \hyperlink{AFM+14}{AFM+14} samples. The [Fe/H] values from \hyperlink{G+18}{G+18} have been transformed into total metallicities in order to be included in this figure, assuming a [Mg/Fe]$\sim$0.1 (the expected abundance following \citealt{Thomas2005} and similar to our mean value, see below). Similarly, the [Fe/H] of \hyperlink{S+09}{S+09} have also been converted to total metallicities, in this case using their published [Mg/Fe]  values.The top panel shows that galaxies further from the cluster centre tend to be younger, largely driven by the \hyperlink{S+09}{S+09} LLGs. These authors reported a spread in ages of 2--10\,Gyr, with the oldest galaxies populating the central regions of the cluster. Our Coma UDGs also reproduce such trend, despite not having a wide distance coverage and not reaching the innermost distances. Interestingly, the \hyperlink{G+18}{G+18} Coma UDGs, located at further distances, behave the opposite way than the outer dwarfs, remaining basically old.

The lower panel shows no trend in metallicity with projected distance, only a mild trend with galaxy type, with UDGs having slightly lower metallicities but similar stellar ages to their co-spatial  \hyperlink{S+09}{S+09} dwarfs. In Paper I we already reported that our UDGs in the cluster core are redder than the \hyperlink{S+09}{S+09} dwarf sample, which can be the result of the harsh cluster environment being more effective at stripping the faint UDGs. This effect could also explain the lower metallicities, having had less time for self-enriching their metallicities (see discussion below). In this panel we find again the different behaviour of the outer UDGs from \hyperlink{G+18}{G+18}, with lower metallicities than their co-spatial dwarfs. In fact, these outer UDGs present a trend where UDGs have lower metallicities as they are further from the cluster centre, also reinforced by the low metallicities of \citet{Kadowaki2017} UDGs at even further distances. Such relatively old ages and lower metallicities of these outer UDGs therefore indicate that these galaxies had an earlier quenching, being examples of the `primordial-type' UDGs. We cannot determine when they were really quenched, but for now we can only point out that this set of UDGs in the outer parts of the Coma cluster may require a different formation path than the UDGs in the region of Coma covered in this paper.

From the results above there is an intriguing fact. It seems like the most central UDGs show properties compatible with the late-infall of dwarf-like type, whereas the outer ones seem to be better described by the primordial-like type. How can we reconcile a late-infall with being already near the cluster core? It can be reconciled by the fact that we are using projected distances and in reality it could be that our central UDGs are at larger 3D distances. This could explain the presence of these UDGs so close to the cluster core, where the number of UDGs is known to decrease significantly (e.g. \citealt{vanderBurg2016}; \citealt{Lee2017}). This would also explain why some of our UDGs have slightly younger ages and lower metallicities than the control LLGs at similar 2D.
\begin{figure}
\centering
\includegraphics[scale=0.52]{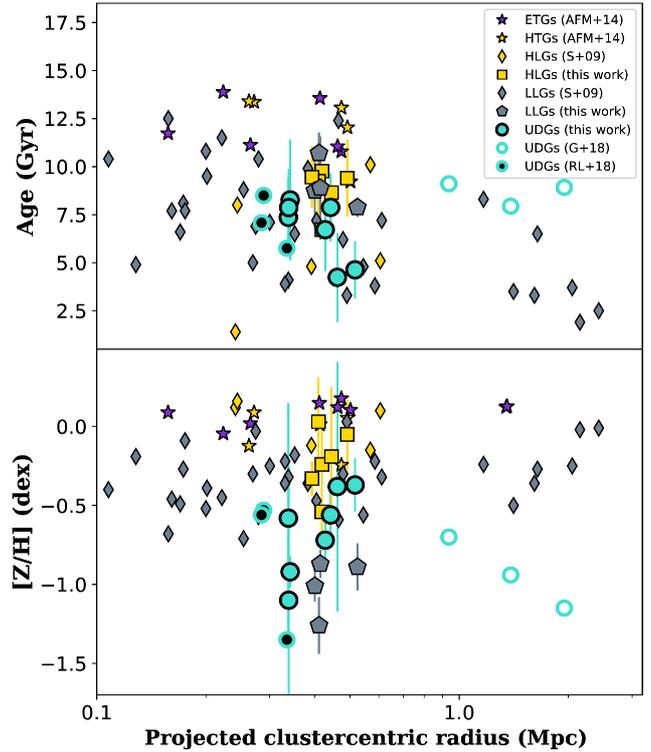}
\label{figure:7}
\vspace{-0.2cm}
\caption{\textbf{Stellar population properties of UDGs with projected cluster-centric radius}. {\it Top:} The mean luminosity-weighted ages of our Coma UDGs are plotted as filled cyan circles, while the G+18 and RL+18 Coma UDGs are shown as open cyan and black-cyan circles, respectively. Our Coma mask LLGs and HLG galaxies are shown as grey pentagons and yellow squares, respectively. We have separated the AFM+14 sample into ETGs (purple stars) and HLG galaxies (yellow stars). The sample of red Coma LLGs from S+09 has also been separated according to a luminosity cut of $R$--band magnitude $R<$\,17, with dwarfs as grey diamonds and normal HLG galaxies as yellow diamonds.The objects in the outskirts tend to show younger mean luminosity-weighted ages, with a drastic variation for the S+09 LLGs. Our Coma UDGs also seem to follow this trend, with the two youngest ones being further from the centre, but such trend is not followed by the outermost UDGs of G+18. {\it Lower panel:} Following the same colour and symbol scheme as above, this panel shows no trend of the total metallicity with cluster-centric radius. However, if we consider inner and outer UDGs, there is a trend of UDGs being more metal poor at larger projected distances.}
\end{figure}

\begin{figure*}
\centering
\includegraphics[scale=0.80]{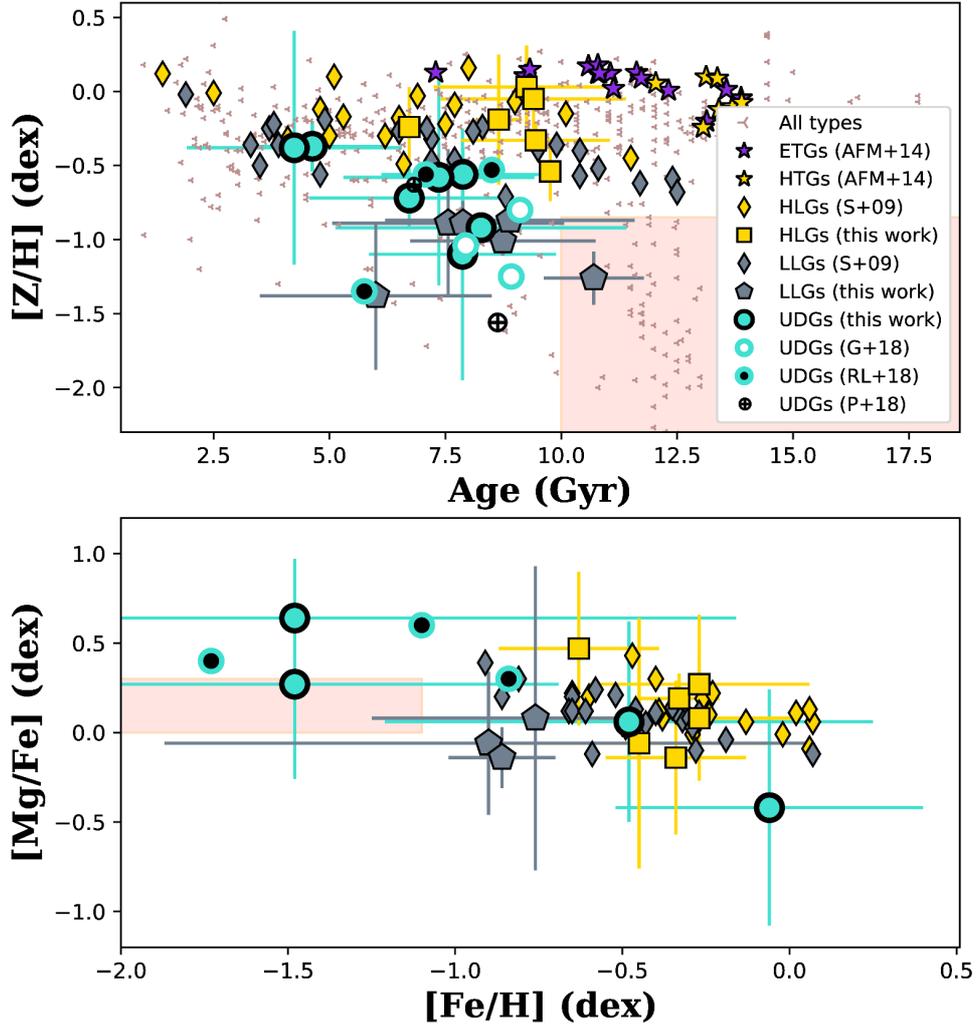}
\label{figure:8}
\vspace{-0.2cm}
\caption{\textbf{Stellar populations of UDGs.} {\it Top panel} shows the age--metallicity distribution of our Coma UDGs, literature ones and the control galaxies from both the observed mask and literature. Colour and symbols scheme is as in Figure 7 but now we also include the Virgo and field UDGs of P+18 (crossed open circles, where the smaller one corresponds to the field DGSAT\,I to show its different environment). The red shaded area corresponds to region covering the properties of disrupted metal-poor globular clusters (the Peng $\&$ Lin (2016) scenario), clearly devoid of UDGs}. Both the LLGs of S+09 and our UDGs show a trend where older galaxies have lower total metallicities.  {\it Lower panel:} [Mg/Fe]--[Fe/H] relation, a proxy for the self-enrichment of a galaxy. Our UDGs with the lowest [Fe/H] values are those with the higher $\alpha$ abundances.
\end{figure*}

\begin{figure*}
\centering
\includegraphics[scale=0.83]{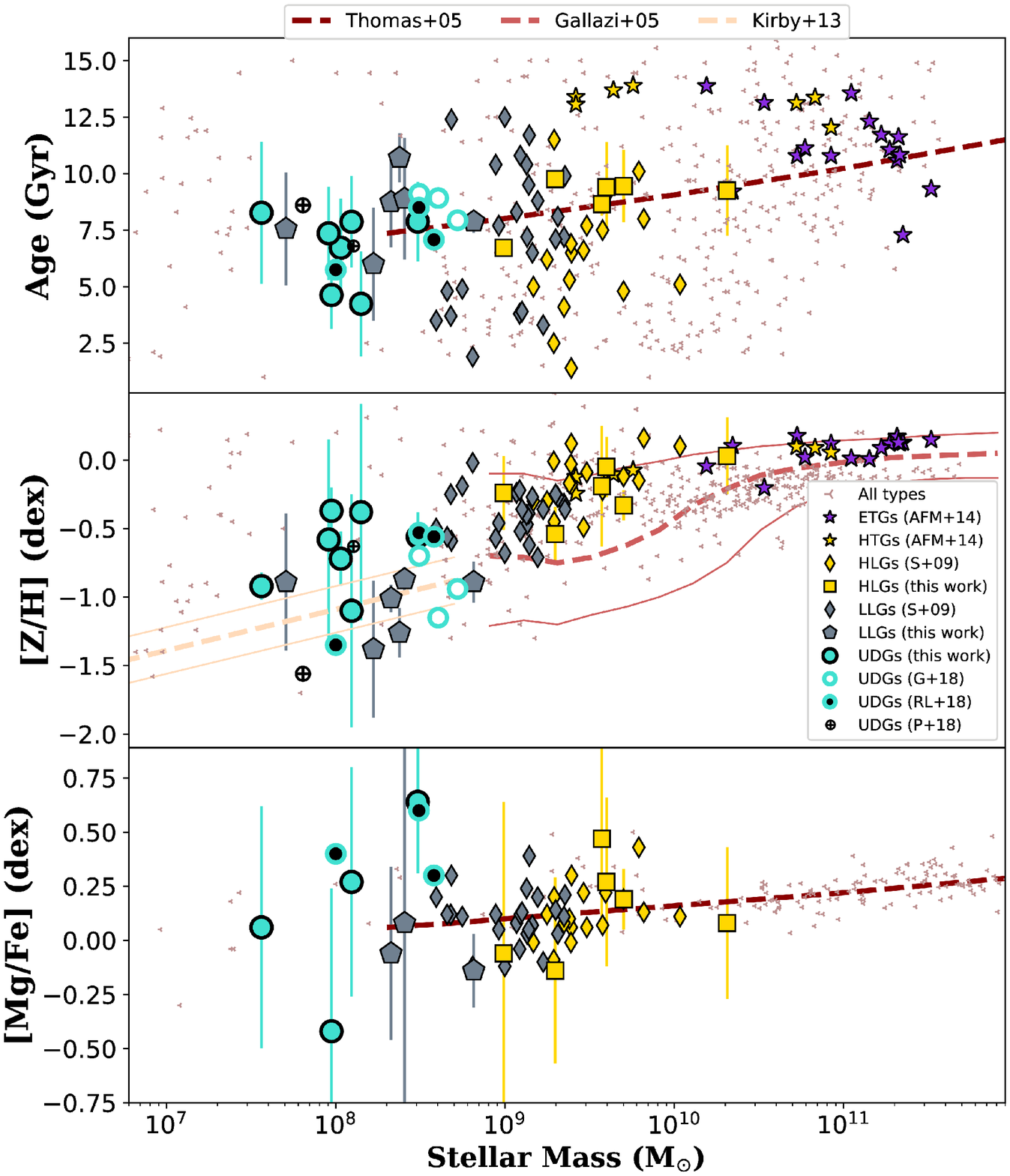}
\label{figure:9}
\vspace{-0.3cm}
\caption{\textbf{Stellar population properties with stellar mass}. {\it Top} panel presents the mass--age relation from \citet{Thomas2005}, which is known to have a high spread as visible by the background red points (all types of literature galaxies from \citealt{Janz2016}). Colours and symbols are as in Figure 5. {\it Middle} panel shows the mass--metallicity relation for both high mass \citep{Gallazzi2005} and low mass galaxies \citep{Kirby2013}. {\it Lower} panel shows the [Mg/Fe]  overabundance ratio with stellar mass relation from \citet{Thomas2005}. In each panel, the UDGs have stellar population properties that are, to first order, consistent with the scaling relations of normal galaxies if they were extrapolated to lower masses, mostly sharing the same properties as the LLGs in Coma (both our Coma mask LLGs and those of S+09).}
\end{figure*}

\subsection{Scaling relations: clues to the origin of UDGs in Coma}
Figure 8 shows the relation between the derived stellar population properties presented in the previous sections for our UDGs and the control galaxies. The top panel shows the age--metallicity plane, with a trend of older UDGs having lower metallicities. The \hyperlink{S+09}{S+09} LLGs also show a similar trend of decreasing metallicity with age, which is not seen neither for the HLGs nor the ETGs. \citet{Peng2016} proposed that the entire stellar content of UDGs are stars originally from disrupted halo globular clusters. Of the two subpopulations found around giant galaxies, dwarf galaxies tend to be dominated by the metal-poor globular cluster subpopulation \citep{Forbes2005}. These globular clusters contain stars that are old (10--13\,Gyr), metal-poor ($-$2.1 $<$ [Z/H] $< -$0.8) and with slightly super-solar $\alpha$--elements (0 $<$ [$\alpha$/Fe] $<$ 0.3; \citealt{Brodie2006}). We have thus highlighted the age--metallicity region expected for such disrupted globular clusters in the top panel of Figure 8. Although one of our UDGs has error bars that could place it in the parameter space of globular clusters, most of our sample UDGs are inconsistent with being such disrupted globular clusters, disfavouring the \citet{Peng2016} scenario.

The lower panel of Figure 8 shows the [Mg/Fe]  ratio \textit{vs} [Fe/H], which is a proxy for the self-enrichment of the galaxies, reflecting the different timescales of nucleosynthesis of the different type of supernovae. It is seen that galaxies having higher [Mg/Fe] ratios, which can be understood as having faster formation timescales \citep{Thomas2005}, are those exhibiting the lowest [Fe/H], compatible with not having time to be enriched in subsequent star formation episodes from more metal-rich gas. This is what is seen for ultra-faint dwarfs (e.g. \citealt{Frebel2014}) and Local Group dwarf and dwarf spheroidal galaxies (e.g. \citealt {Venn2004}; \citealt{Kirby2011}; \citealt{Revaz2018}), with [Fe/H]-$\alpha$ ratios similar to the Milky-Way field stars, whereas some of our UDGs (the most enhanced ones) resemble more to the Milky-Way halo stars (e.g. \citealt{Venn2004}).

The final step in the analysis is to see how the stellar populations of our UDGs correlate with their stellar mass and to find out if their location in such relations can shed any further light into the origin(s) of UDGs. In Figure 9 we show the age, metallicity and $\alpha$--element abundance ratio against stellar mass for both Coma and other literature UDG. The control galaxies used in the previous sections (LLGs and HLGs contained in our DEIMOS mask, HLGs and ETGs from \hyperlink{AFM+14}{AFM+14} and the Coma LLGs and HLG galaxies from  \hyperlink{S+09}{S+09}) are also included. The stellar masses for the \hyperlink{S+09}{S+09} sample have been converted from the luminosity, $r$--band magnitude and the M/L ratio corresponding to their SSP age and metallicity, consistent with the way we measured the stellar masses in Section 3 for our DEIMOS objects, using the {\tt MILES} SSP models. Before proceeding we remind the reader the caveat that all galaxies are not from the same environment/location within the cluster. However, as seen in Figure 7 only a few of the  \hyperlink{S+09}{S+09} LLGs are located at further distances and we highlight that both UDGs from \hyperlink{P+18}{P+18} belong to either another cluster or the field. 

The top panel includes the mass--age relation for massive systems from \citet{Thomas2005}. The stellar ages of UDGs are slightly lower than the outer UDGs of \hyperlink{G+18}{G+18}, but more similar to the \hyperlink{RL+18}{RL+18} and the \hyperlink{P+18}{P+18} (both cluster and field galaxies) ones. Also the the dwarf galaxies in our DEIMOS mask share the same ages. The HLGs in the mask show a wider spread in ages, similar to the one seen for the  \hyperlink{S+09}{S+09} sample of red dwarf galaxies. This panel shows that the ages of the UDGs are consistent with the mass--age relation of massive systems if such a relation was extrapolated to lower masses.  

The middle panel of Fig. 9 shows the well-known mass--metallicity relation, represented by \citet{Gallazzi2005} for high mass galaxies and by \citet{Kirby2013} for low-mass ones. Typically, our UDGs lie above the mass--metallicity relation expected for low-mass systems of \citet{Kirby2013}. In fact, our UDGs seem to rather follow the low-mass end of the \citet{Gallazzi2005} relation, following the tail described by the Coma LLGs from \hyperlink{S+09}{S+09}, but presenting slightly lower metallicities than the latter. \hyperlink{S+09}{S+09} already reported that such galaxies seemed to follow the mass--metallicity relation of more massive galaxies rather than the lower mass systems of \citealt{Kirby2013}, in agreement with our findings for the UDGs and some dwarfs in our mask. The UDG that strongly departs from the rest is the Virgo UDG VCC\,1287, which has similar old ages but lower metallicity (\hyperlink{P+18}{P+18}). 

The lower panel of Fig. 9 shows the relation of the stellar mass with [Mg/Fe]. Our UDGs have a mean value of [Mg/Fe] $\sim$\,0.13\,dex, which is compatible with the relation for normal galaxies from \citet{Thomas2005} extrapolated to lower stellar masses. Such abundances are similar to those found for other low-mass galaxies and compact stellar systems such as ultra compact dwarfs and compact ellipticals (e.g. \citealt{Janz2016}; \citealt{Ferre-Mateu2018}). This would support the claim that some compact stellar systems could be the remnant nuclei of clumpy UDGs that have the bulk of their stellar content stripped during their journey across the cluster environments \citep{Janssens2017}. However, if we include the values from \hyperlink{RL+18}{RL+18}, the mean abundance value increases to [Mg/Fe]$\sim$\,0.3. Both our UDGs with super-solar abundances are those with the lowest values of [Fe/H], which can be understood in terms of a strong suppression of the Fe rather than an over-abundance of Mg. This could be related to these galaxies being formed on shorter timescales. Although we were not able to obtain the abundances for all the UDGs, we can see that the only UDG with an $\alpha$--abundance lower than the rest (Yagi275) is the one with young ages, showing an extended SFHs with a large contribution of recent star formation. This topic will be further discussed in Mart\'in-Navarro et al. (submitted). \\

We have thus seen that in general, the properties of our UDGs are similar to the LLGs observed in our mask and those from \hyperlink{S+09}{S+09} located at similar projected cluster-centric radii. This is further reinforced by the recent results of \hyperlink{RL+18}{RL+18} for UDGs in the same neighbourhood. Although none of the properties alone is enough to discriminate between the proposed origins, when adding all the indicators together, a more clear picture is revealed. Our findings support the idea that the UDGs studied in this work share a common dwarf-like origin, where they were created outside the cluster environment as dwarfs. While our data cannot discriminate whether these UDGs have extended sizes due to internally driven-processes (i.e. outflow-driven gas feedback \citealt{DiCintio2017}) or simply because they lived in high-spin haloes that prevented them from condensing (e.g. \citealt{Amorisco2016}; \citealt{Rong2017}), our results support a scenario where our UDGs were quenched recently due to a late infall into the Coma cluster.

Such a `dwarf-like' origin is consistent for the UDGs populating the central region of Coma covered in this work and those from \hyperlink{RL+18}{RL+18}. But it does not necessarily extend for all UDGs in Coma. While our interpretation seems to reinforce the dwarf-like nature for the Virgo UDG VCC\,1278 \citep{Beasley2016} and DF17 (e.g. \citealt{Peng2016}; \citealt{BeasleyT2016}), the properties of other Coma UDGs at larger cluster-centric radii in Coma (e.g. DF44, DF07 and the \citealt{Kadowaki2017} UDGs) seem to suggest a different path of formation. This was also indicated in Paper I, where a few UDGs were shown to be compatible with being quenched earlier, hence proposed as `primordial' galaxies.

Therefore, not having yet determined the relative proportions of dwarf-like and normal galaxy-like UDGs in Coma, leaves the enigma of the nature of the UDG population still open to debate. Future work should be able to tackle this issue by compiling a large, statistical sample of UDGs at different cluster-centric radii with sufficiently high S/N to alleviate the several caveats we encountered during the analysis. This will help put further constraints on the (diverse) origin(s) of UDGs and their relative importance in cosmological models.

\section{Conclusions}
We have presented a new spectroscopic study of the stellar content of seven UDGs nearby the core of the Coma cluster, the largest to date, in order to further investigate the possible origins of UDGs and their overall nature. The analysis of their stellar ages, metallicities, [Mg/Fe]  abundances and SFHs, combined with similar data in the literature, indicates that the UDGs in our sample were accreted into the Coma cluster later than other luminous galaxies and normal dwarfs, being compatible with a rather recent infall. Our UDGs have intermediate ages ($\sim$7\,Gyr), low metallicities ([Z/H]\,$\sim\,-$\,0.7) and slightly super-solar $\alpha$--abundances ([Mg/Fe] \,$\sim$\,0.13). We find that in general, all their stellar population properties are consistent with the low surface brightness dwarf galaxies in our sample and other co--spatial dwarfs, and are inconsistent with the suggestion of \citet{Peng2016} that their stellar component is due entirely to disrupted globular clusters. In fact, UDGs seem to follow most of the higher-mass galaxies scaling relations (i.e. a continuation of mass--age, mass--metallicity and mass--$\alpha$ elements), further supporting their dwarf-like origin. 

The mean ages obtained from the different approaches used in this work are also consistent with the predicted age distributions of UDGs from the cosmological simulations of \citet{Chan2018} and \citet{Rong2017}, with formation redshifts of $z\,<$\,1. Their sustained star formation rates down to recent times further support the assumption of a late infall into the cluster core, with `quenching' times of $\sim$11\,Gyr. These results from the stellar populations are further reinforced by the results in Paper I of the series, where recent infall into the cluster has also been found for most of our Coma UDGs. All this disfavours the scenarios whereby early-forming primordial galaxies failed to evolve and rather supports the dwarf-like origin for this sample of Coma UDGs.

To summarise, our stellar population results for seven UDGs in the Coma Cluster contribute to the growing evidence towards a dwarf-like origin for many UDGs in the literature. However, the properties of other studied UDGs invoke the need for other formation pathways, leaving the enigma of the nature of UDGs as an open question. 

\section*{Acknowledgements}
The authors thank the constructive comments from the referee, which have improved the clarity of the manuscript. We would like to thank P. van Dokkum and C. Conroy for insightful discussions. AFM acknowledges the NIHAO collaboration (P.I. Andrea Macci\'o) for sharing unpublished data on their simulated UDGs and thanks Arianna di Cintio for the effort of getting such data together. AFM and DAF acknowledge the ARC for financial support via DP160101608. AJR was supported by NSF grant AST-1616710 and as a Research Corporation for Science Advancement Cottrell Scholar. JB, BA and IMN were supported by NSF grant AST-1616598. SB was supported by the AAO PhD top-up Scholarship. M.B.S. acknowledges financial support from the Academy of Finland, grant 311438.\\
The data presented herein were obtained at the W. M. Keck Observatory, which is operated as a scientific partnership among the California Institute of Technology, the University of California and the National Aeronautics and Space Administration. The Observatory was made possible by the generous financial support of the W. M. Keck Foundation. The authors wish to recognise and acknowledge the very significant cultural role and reverence that the summit of Maunakea has always had within the indigenous Hawaiian community.  We are most fortunate to have the opportunity to conduct observations from this mountain. \textit{M\=alama ka '\=aina}.


\bibliography{coma_udgs}
\bibliographystyle{mn2e}

\appendix

\begin{figure*}
\centering
\includegraphics[scale=0.55]{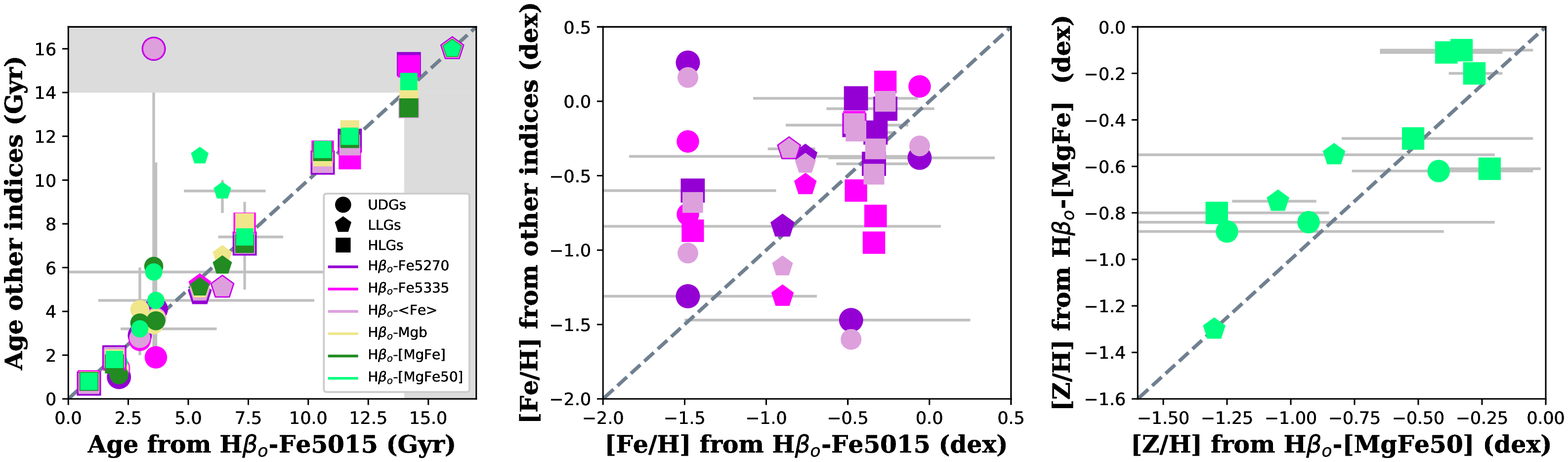}
\label{figure:A1}
\vspace{-0.2cm}
\caption{\textbf{Line index results.} We show the resulting age, [Fe/H] and total metallicity [Z/H] from different pairs of indices, using the age sensitive H$\beta_{\rm o}$ for all the galaxies in the mask with enough S/N. Shaded areas in the age panel correspond to extrapolated measurements, where the indices fall outside of the model grid. The derived ages are very robust across the different indices, but larger differences are found in the metallicity panels, due to some poor line measurements.}
\end{figure*}

\begin{figure*}
\centering
\includegraphics[scale=0.7]{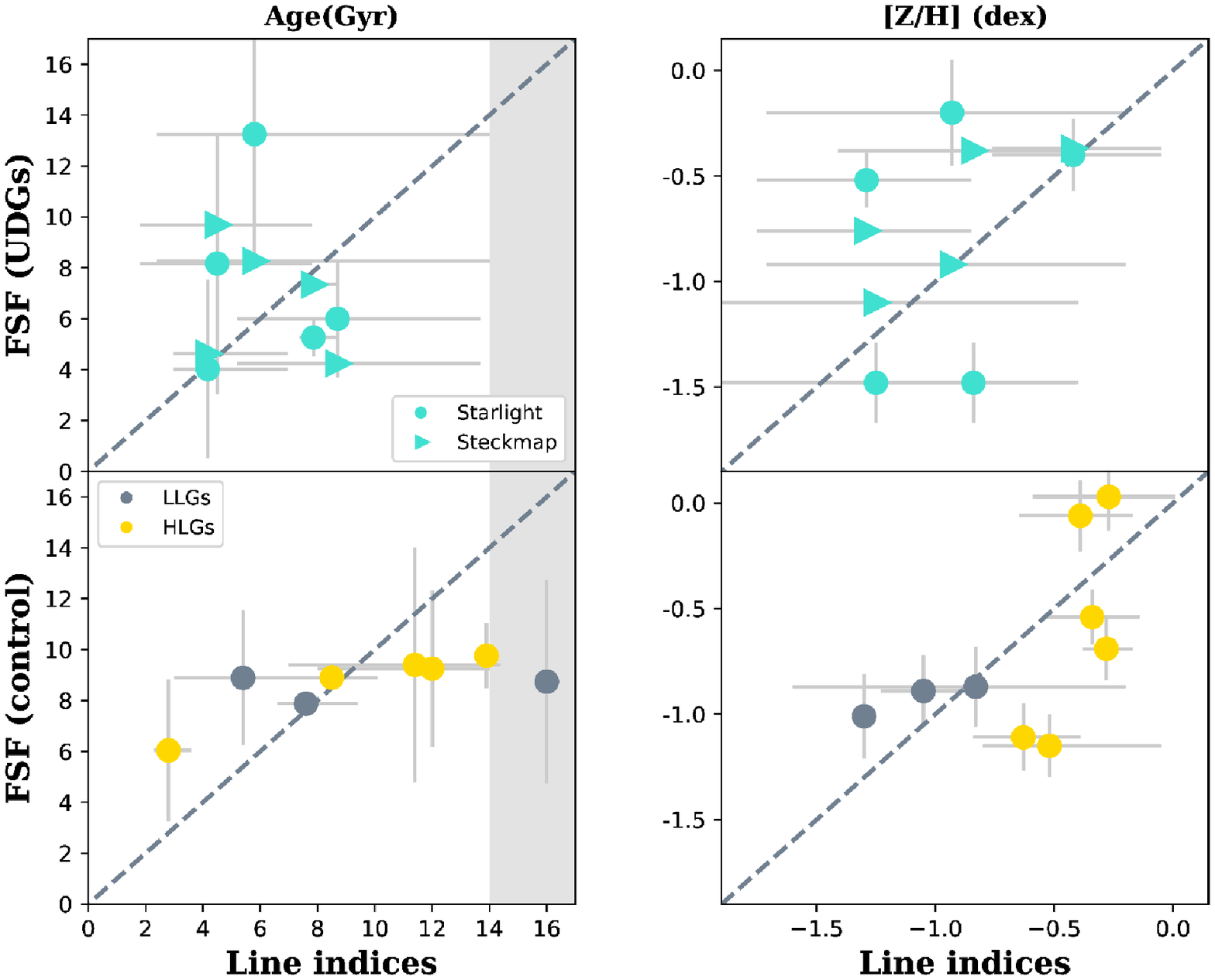}
\label{figure:A2}
\vspace{-0.2cm}
\caption{\textbf{Comparison of the methodologies.} We show the comparison between the results obtained by using the full spectral fitting approach (FSF) and those from the line indices. \textit{Top} panels show the age (\textit{left}) and the total metallicity (\textit{right}) for the Coma UDGs. \textit{Bottom} panels show the same properties derived with {\tt STARLIGHT }for the control galaxies in our study, i.e Coma cluster HLG and dwarfs. Shaded areas in the age panel correspond to extrapolated measurements, where the indices fall outside of the model grid.}
\end{figure*}

\section{Stellar population analysis}
Here we present in detail the stellar populations derived from all the different index-index pairs and both full spectral fitting codes. We measure the most relevant indices in our spectral region, which are H$\beta$ and H$\beta_{\rm o}$ \citep{Cervantes2009} as age-sensitive indices, and Fe5015, Mgb$_{5177}$, Fe5270 and Fe5335 as the metallicity-sensitive ones. We also use the following combined indices <Fe>$^{\prime}$ =\,(0.72 $\times$ Fe5270 + 0.28 $\times$ Fe5335); [Mg/Fe] $^{\prime}$= $\sqrt{\mathrm{Mgb \,\,\times <Fe>^{\prime}}}$ and [MgFe50] =\,(0.69$\times$Mgb + Fe5015)/2 (\citealt{Thomas2003}; \citealt{Kuntschner2010}). 

For the rest of this section we use the highly age sensitive H$\beta_{\rm o}$ index rather than the H$\beta$ one, as it is shown to provide more orthogonal model grids. From each pair of age-metal indices we thus derive an SSP age and metallicity, as shown in Figure A1. In the age panel (left), the shaded areas correspond to extrapolated measurements, where the indices fall outside of the model grid. Any galaxy in that area should thus be considered as very old. We show the [Fe/H] metallicities from all the different Fe lines in the middle panel. For the total metallicity [Z/H] (right) we use the combination of the age sensitive index with either of the combined [MgFe] indices, known to be insensitive to $\alpha$-enhancements. Overall Figure A1 shows that the SSP ages derived with the different sets of indices are very robust. However, it is more difficult to obtain robust values for [Fe/H] and [Z/H] as the lines are in many cases affected by noise in the spectra or are near the edge of the spectra (in particular for Fe5335 and thus the combined $\mathrm{<Fe>^{\prime}}$).

We then obtain the [Mg/Fe]  ratios using the same approach as in \citet{Vazdekis2015}. We use the metallicity estimates $Z_{\rm Mgb}$ and $Z_{\rm Fe5015}$ obtained previously and the proxy [$Z_{\rm Mg}/Z_{\rm Fe5015}$] = $Z_{\rm Mgb}\,-\,Z_{\rm Fe5015}$. Then, using the empirical relation of \citet{Vazdekis2016} we transform it, using [Mg/Fe]  = 0.59 $\times$ [$Z_{\rm Mg}/Z_{\rm Fe5015}$]. 

We finally perform the full spectral fitting approach to obtain mean luminosity- and mass-weighted stellar population properties. In addition, we recover the SFH of each galaxy, which is the amount of stellar mass that is created at a given time. We use both {\tt STARLIGHT} \citep{CidFernandes2005}) and {\tt STECKMAP} \citep{Ocvirk2006} for this analysis. We remind the reader that {\tt STECKMAP} allows the use of non-flux calibrated spectra, whereas {\tt STARLIGHT} demands a relative-flux calibration (which contributes to a smaller sample of usable UDGs, i.e. we omit Yagi392 and Yagi398). 

In Figure A2 we show a comparison of age and total metallicity derived from full spectral fitting with those from line indices for Coma UDGs (top) and HLG plus dwarf galaxies (bottom). This figure shows that the ages are in general well constrained across the different methods. In some cases, larger  differences are seen but they can be understood when carefully inspecting the spectra (e.g. H$\beta$ lines not well fitted). The variations in the total metallicity are slightly larger, but they are all compatible with having low metallicities. Having potentially three different stellar population results to hand, we choose for the analysis those values that we consider to be most robust, i.e. taking into account issues with crucial line indices or issues with a bad flux calibration. The final values are those used throughout Section 4 and are quoted in Table 2.


\bsp	
\label{lastpage}
\end{document}